\documentclass[twocolumn,manuscript]{emulateapj}
\usepackage{graphicx}
\usepackage{amssymb}
\usepackage{natbib}
\usepackage{hyperref}
\def \eg                {\hbox{e.g.,~}}

\def \etal              {\hbox{et~al.~}}

\def \arcsec            {{^{\prime\prime}}}

\def \kpc               {{\rm\ kpc}}

\def \H0                {{\rm\ H_{0}}}

\def \mum               {{\rm\ {\mu}m}}
\def \kel               {{\rm\ K}}

\begin{document}
  \received{2009 January 7}
  \accepted{2009 September 3}
  \slugcomment{To appear in ApJ November 2009 Issue 1}

  \shortauthors{Deo \etal}
  
  \shorttitle{The Mid-Infrared Continua of Seyfert Galaxies}
  
  \title{The Mid-Infrared Continua of Seyfert Galaxies}
  
  \author{Rajesh. P. Deo\altaffilmark{1}, Gordon. T. Richards\altaffilmark{1},
    D. M. Crenshaw\altaffilmark{2}, S. B. Kraemer\altaffilmark{3}}
  
  \altaffiltext{1}{Department of Physics, Drexel University, 3141, Chestnut
    Street, Philadelphia, PA 19104-2816, USA; rpd@physics.drexel.edu and
    gtr@physics.drexel.edu}

  \altaffiltext{2}{Department of Physics and Astronomy, Georgia State
    University, Atlanta, GA 30303, USA; crenshaw@chara.gsu.edu}

  \altaffiltext{3}{Catholic University of America, and the Exploration of the
    Universe Division, NASA's Goddard Space Flight Center, Code 667,
    Greenbelt, MD 20771, USA; kraemer@yancey.gsfc.nasa.gov}
  
  \begin{abstract}
    An analysis of archival mid-infrared (mid-IR) spectra of Seyfert galaxies
    from the \textit{Spitzer Space Telescope} observations is presented. We
    characterize the nature of the mid-IR active nuclear continuum by
    subtracting a template starburst spectrum from the Seyfert spectra. The
    long wavelength part of the spectrum contains a strong contribution from
    the starburst-heated cool dust; this is used to effectively separate
    starburst-dominated Seyferts from those dominated by the active nuclear
    continuum. Within the latter category, the strength of the active nuclear
    continuum drops rapidly beyond $\sim 20\mum$. On average, type 2 Seyferts
    have weaker short-wavelength active nuclear continua as compared to type 1
    Seyferts. Type 2 Seyferts can be divided into two types, those with strong
    poly-cyclic aromatic hydrocarbon (PAH) bands and those without. The latter
    type show polarized broad emission lines in their optical spectra. The
    PAH-dominated type 2 Seyferts and Seyfert 1.8/1.9s show very similar
    mid-IR spectra. However, after the subtraction of the starburst component,
    there is a striking similarity in the active nuclear continuum of all
    Seyfert optical types. PAH-dominated Seyfert 2s and Seyfert 1.8/1.9s tend
    to show weak active nuclear continua in general. A few type 2 Seyferts
    with weak/absent PAH bands show a bump in the spectrum between 15 and
    20$\mum$. We suggest that this bump is the peak of a warm
    ($\sim$200$\kel$) blackbody dust emission, which becomes clearly visible
    when the short-wavelength continuum is weaker. This warm blackbody
    emission is also observed in other Seyfert optical sub-types, suggesting a
    common origin in these active galactic nuclei.
  \end{abstract}
  
  \section{Introduction}
  The complete mid-infrared (mid-IR) spectra of Seyfert galaxies \citep[a
  class of active galactic nuclei (AGN):][]{1943ApJ....97...28S} have been
  available only in recent past
  \citep[\eg][]{2002A&A...393..821S,2005SSRv..119..355V,2005ApJ...633..706W,2007ApJ...671..124D}.
  Many {\it Spitzer} spectra of nearby Seyfert galaxies show a strong
  contribution from star-forming features in the form of poly-cyclic aromatic
  hydrocarbon (PAH) bands
  \citep[\eg][]{2000A&A...357..839C,2006AJ....132..401B,2008ApJ...676..836T}.
  A number of these features, such as the 7.7$\mum$ and the 17$\mum$ PAH
  complex, are strongly blended with each other, complicating the estimation
  of the underlying active nuclear continuum. Further, the mid-IR opacity
  includes a strong contribution from the silicate bands at $10$ and $18\mum$.
  Thus, estimating the intrinsic active nuclear continua in the mid-IR is a
  non-trivial task, with the primary hurdle being the subtraction of the
  starburst component.
  
  Previous studies have hinted that the continuum in the 1--8$\mum$ range is
  non-stellar in origin and is likely a result of thermal emission from dust
  heated close to sublimation temperature by the optical/ultra-violet
  continuum from the central source
  \citep{1986ApJ...308...59E,2001AJ....121.1369A,2004ApJ...614..122I,2008arXiv0807.4695M}.
  In a multi-wavelength photometric study of spectral energy distributions
  (SED) of predominantly radio-quiet Sloan Digital Sky Survey (SDSS) quasars,
  \citet{2006ApJS..166..470R} and \citet{2007ApJ...661...30G} noted that the
  1--8$\mum$ spectral index ($\alpha_{\nu}$) is strongly anti-correlated with
  infrared luminosity in type 1 quasars. The more luminous quasars have
  flatter 1--8$\mum$ slopes. A linear correlation between the optical
  continuum luminosity and the infrared luminosity suggests that the observed
  bump around $\sim2.2\mum$ in the SED is driven by the dust re-emission. For
  example, the $2.2\mum$ bump is clearly visible in the near-IR spectrum of
  Mrk~1239 \citep{2006MNRAS.367L..57R}.
  
  Due to the presence of strong PAH bands in the 5--8$\mum$ range in
  many Seyfert spectra, direct measurement of the active nuclear continuum in
  this region is only possible in sources with very weak or absent PAH
  features. An alternative is to subtract the starburst contribution using a
  template starburst spectrum and then study the residual continuum. We take
  this simple approach in this paper. In Section~2, we describe our archival sample
  and the data analysis techniques. We discuss the observed mid-IR spectra in
  Section~3. In Section~4, we use simple continuum diagnostics that allows us to classify
  Seyfert mid-IR spectra into PAH-dominated and AGN-dominated groups and
  understand continuum properties. In Section~5, we discuss the starburst
  contribution in Seyfert spectra and the resulting continuum shapes after
  subtraction of the starburst template spectrum. In Section~6, we summarize our
  results.
  
  \section{Sample Selection and Data Analysis}
  
  We consider a sample of Seyfert galaxies derived from the {\it Spitzer Space
    Telescope} archives. This sample is listed in Table~\ref{tab:objdata} with
  redshifts obtained from the NASA/IPAC Extragalactic
  Database\footnote{http://nedwww.ipac.caltech.edu/} (NED) along with the {\it
    Spitzer} archive numbers (AORKEY). We primarily use archival InfraRed
  Spectrograph \citep[IRS;][]{2004ApJS..154...18H} spectra extracted from
  programs 3069 (PI: J. Gallimore, mapping mode spectra\footnote{See:
    http://ssc.spitzer.caltech.edu/documents/SOM/}), 3374 (PI: S. Kraemer,
  staring mode spectra) and \citet[][; Mrk 3, low-resolution staring mode
  spectrum]{2005ApJ...633..706W}. \citet{2007ApJ...671..124D} presented single
  slit extractions of spectra from above mentioned programs for about half of
  the current sample. During the initial work on that paper, CUBISM software
  was not yet available \citep{2007PASP..119.1133S}. In this paper, we expand
  that sample and consider the complete IRS mapping-mode datasets from program
  3069 along with other archival datasets now available. This sample was
  chosen due to its large size and good sampling of different Seyfert optical
  types. We also test if the use of complete mapping-mode spectra leads to
  changes in the short wavelength continuum. After the initial analysis, we
  noted that our sample lacked type 2 Seyferts with detected polarized broad
  emission lines. So, we added 16 Seyfert 2s selected from the compilation of
  \citet{2003ApJ...583..632T} for which low resolution IRS spectra are
  available from the archive. These objects come from a number of different
  programs and their AOR numbers are listed in Table~\ref{tab:objdata}.
  Further, we obtained optical Seyfert classifications from the compilation of
  \citet[][VCV, hereafter]{2006A&A...455..773V} and host galaxy minor-to-major
  axis ratios ($b/a$) from NED. These are listed in Table~\ref{tab:objdata}.
  This paper contains a sample of 109 Seyfert galaxies with low-resolution IRS
  spectra.
  
  The mapping-mode observations were reduced as below. From the low-resolution
  basic calibrated data (BCD) spectral images (for short-low, SL and long-low,
  LL modules), we built three-dimensional spectral cubes (two spatial
  ($x$,$y$) and one wavelength ($z$) dimension) using the CUBISM software
  package \citep{2007PASP..119.1133S}. We wrote Interactive Data Language
  (IDL) routines to script internal CUBISM routines to automate the reduction
  of spectral cubes. The pixel backtracking facility available within CUBISM
  was used to clean the cubes of bad pixels. One-dimensional spectra were
  extracted from these cubes using a common rectangular aperture for all four
  modules. These extraction apertures (see Table~\ref{tab:objdata}, Column 9)
  were selected to encompass as small a region as possible in the LL1 module
  without introducing pixel-aliasing effects in the extracted one-dimensional
  LL1 spectra. We note an increase in the absolute SL flux density values as
  compared to single slit extractions. This is expected due to the different
  pixel scales of the SL and LL modules and the larger aperture area as
  compared to a single slit. The extracted one-dimensional spectra were then
  imported into the SMART software package \citep{2004PASP..116..975H} to
  normalize the separate modules. In most cases no multiplicative shifts were
  necessary due to matched aperture extractions. When corrections were
  necessary, the correction factor was always less than 10\%. The ends of the
  spectral orders were clipped to remove data points affected by the reduced
  spectral response. The spectra were then re-binned to a common wavelength
  grid.
  
  The staring-mode spectra were extracted as below. We started with BCD
  products obtained from the {\it Spitzer} archive. We median-combined
  multiple data collection event (DCE) image files into one image per module,
  order and nod-position. We differenced images from opposite orders to remove
  the sky background and then extracted spectra using the tapered-column
  extraction option within SMART. For cases, when differencing opposite orders
  was not possible, we used opposite nod positions. All the image and spectrum
  extraction operations were carried out inside SMART. The extracted spectra
  were cleaned by removing deviant data points and normalizing the spectra
  from different modules to form the complete mid-IR spectrum.
  Figure~\ref{fig:all_spec} shows the complete mid-IR spectra grouped
  according to their Seyfert types from Table~\ref{tab:objdata}.
  
  We measured the mid-IR continuum at $5.5$, $10$, $14.7$, $20$ and
  $30\mum$, averaging within a window of $1\mum$, on the rest-frame spectra.
  These specific spectral regions where chosen to minimize the contribution
  from emission lines or PAH bands, thus primarily sampling the mid-IR
  continuum. We also measured the optical depth at $9.7\mum$. The silicate
  strength commonly quoted in {\it Spitzer} studies is $S_{9.7} = \ln
  (f_{\lambda}/f_{C})$, which translates to $- \tau_{9.7}$ in our case. We
  follow the convention that negative optical depth implies a silicate
  emission feature. The peaks of these emission features typically occur
  around $10.5\mum$. We present optical depth as measured at $9.7\mum$ for
  such sources. The continuum and optical depth measurements are given in
  Table~\ref{tab:contflux}. We also measured the equivalent widths and fluxes
  of PAH bands and the narrow emission lines. In this paper, we restrict
  ourselves only to continuum and optical depth measurements.
  
  In all the figures ahead, Seyfert 1s (including 1.2s and 1.5s) are
  represented by filled circles (types S1, S1.0, S1.2, and S1.5 from VCV).
  Seyfert 2s with broad emission lines detected in their polarized optical
  spectra \citep[type S1h in VCV; and called hidden broad-line region, HBLR,
  in][]{2003ApJ...583..632T} are represented by filled diamond symbols.
  Seyfert 2s with undetected broad emission lines in their polarized optical
  spectra \citep[type S2 from VCV, and called non-HBLR
  in][]{2003ApJ...583..632T} are represented with open squares. Seyfert
  1.8/1.9s are represented with open triangles (S1.8 and S1.9 from VCV).
  Galaxies classified as LINER (type S3 from VCV) are represented by cross
  symbols. Each figure includes legends describing all Seyfert optical
  sub-types mentioned here.

  \section{The Mid-IR Spectra}
\begin{figure}[!ht]
    \begin{center}
      \includegraphics[width=\textwidth]{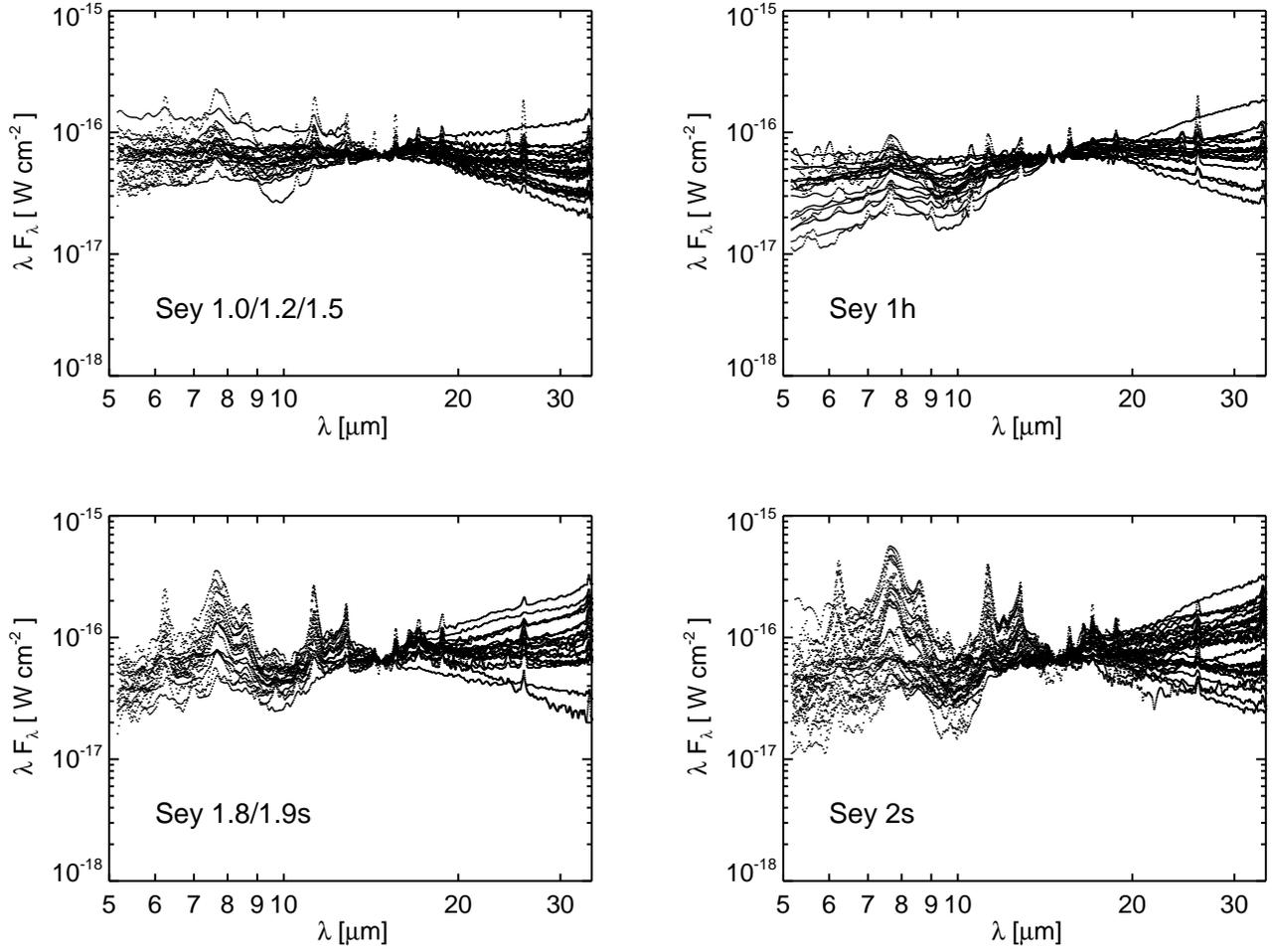}
    \end{center}
    \caption{Mid-infrared spectra of the entire sample normalized at
      14.7$\mum$. Top left: Seyfert 1s; top right: Seyfert 2s with evidence of
      broad optical emission lines in polarized light; bottom left: Seyfert
      1.8/1.9s spectra dominated by PAH features; bottom right: Seyfert 2
      spectra dominated by PAH features. The short-wavelength continuum of
      PAH-weak type 2 Seyferts (top right) is on average weaker and steeper
      than type 1 Seyferts (top left). Seyfert 1.8/1.9s and PAH-strong type 2
      Seyferts (bottom right) show very similar spectra.}
    \label{fig:all_spec}
  \end{figure}

  Complete mid-IR spectra of the sample are presented in
  Figure~\ref{fig:all_spec}. The top left panel shows type 1 Seyfert spectra.
  The top right panel shows type 2 Seyferts with broad-line region (BLR)
  detected in polarized light
  \citep[\eg][]{1985ApJ...297..621A,2003ApJ...583..632T}. The bottom left
  panel shows Seyfert 1.8/1.9s; these spectra are dominated by
  starburst-related PAH emission. These nuclei are likely weak emitters with a
  dominant contribution from circum-nuclear starbursts
  \citep{2007ApJ...671..124D}. The bottom right panel shows type 2 Seyferts
  with dominant starburst contribution (strong PAH bands). Clearly the mid-IR
  spectra of Seyfert 1.8/1.9s are very similar to these type 2 Seyferts.
  Polarized broad emission lines have not yet been detected in these
  PAH-dominated type 2 Seyferts. This suggests that the starburst contribution
  from the host galaxy dominates over the active nuclear continuum in these
  sources. One of the goals of this study is to understand if these
  starburst-dominated nuclei have similar mid-IR continuum properties as
  starburst-weak type 1 and type 2 nuclei.
  
  Examination of the spectra in Figure~\ref{fig:all_spec} shows that on
  average, the continuum is similar between all four classes. PAH-weak type 2
  spectra (top right) show weaker and steeper short-wavelength continuum than
  PAH-weak type 1 spectra (top left) which are flatter. This is a direct
  evidence that short-wavelength mid-IR emission is absorbed to a certain
  degree in type 2 objects by the intervening dust torus.

  Previous research work
  \citep[\eg][]{1986ApJ...308...59E,1996A&A...315L.129R,2001A&A...379..823K,2006ApJ...649...79S}
  has suggested that the mid-IR SED is likely composed of three thermal
  components: hot ($\sim 1000\kel$), warm ($\sim 200\kel$) and cold ($\sim
  60\kel$) dust emissions. This is reflected in the {\it IRAS} 25 and 60$\mum$
  photometry with the active nuclear SED being ``warm'' compared to the
  ``cool'' starbursts.

  Almost all spectra show a curious emission peak/bump between 15 and
  20$\mum$. This emission should not be confused with the silicate $18\mum$
  peak (which is always much weaker than the 10$\mum$ peak) or the $17\mum$
  PAH complex. The feature is broad and the peak wavelength varies between 15
  and 20$\mum$. This bump is observed most clearly in the top right panel of
  Figure~\ref{fig:all_spec} (S1h sources), as the short-wavelength continuum
  is weak in these sources. The presence of this feature in many other spectra
  (some with strong PAH contributions also) suggests a common origin in most
  Seyfert sub-types.

  In Figure~\ref{fig:bump}, we subtract the spectrum of Mrk 335 (a Seyfert
  1.2) from the starburst-subtracted spectrum\footnote{See Section~5 ahead for
    details of the starburst subtraction process.} of Mrk 766 (a Seyfert 1.5).
  The Mrk 335 spectrum is a power-law-like spectrum with no hint of the
  15--20$\mum$ bump. The residual Mrk 766 spectrum shows the 15--20$\mum$
  bump. We compare the residual spectrum to the Mrk 3 (type 2, S1h) spectrum.
  Both spectra are almost identical. This suggests that the 15--20$\mum$ bump
  is only visible when the power-law-like hot dust component is being
  absorbed, as in type 2 sources. Thus, it seems that the continuum shape of
  the mid-IR AGN spectrum depends on which of the three hot, warm, and cold
  components is brighter than others. In type 1 spectra, the hot component
  dominates; in type 2 spectra, the warm component dominates; while
  starburst-dominated objects show an excess of cold dust emission.

  \section{The Observed Mid-IR Continuum}
  Figure~\ref{fig:continua} shows a plot of spectral indices for all galaxies
  in the sample. Here, the spectral index\footnote{The wavelength index
    $\alpha_{\lambda}$ and the frequency index $\alpha_{\nu}$ are related by
    $\alpha_{\nu} = - (\alpha_{\lambda} + 2)$.} is defined as
  \begin{equation}
    \alpha_{\lambda_{1} - \lambda_{2}} = \log_{10}(f_{\lambda_{1}}/f_{\lambda_{2}})/\log_{10}(\lambda_{1}/\lambda_{2})
  \end{equation}
  where $f_{\lambda_{1}}$ is the flux density in ${\rm W}\,{\rm
    cm}^{-2}\,\mum^{-1}$ and wavelengths are in $\mum$. The dotted lines at
  $\alpha_{\lambda}(5.5\textrm{--}14.7) = -0.5$ and
  $\alpha_{\lambda}(20\textrm{--}30) = -0.5$ divide the figure in four
  quadrants. Three important conclusions can be drawn from this comparison:
  (1) most HBLR Seyfert 2s (class S1h in Table~\ref{tab:objdata}) are in the
  left quadrants; (2) most Seyfert 1s are in the bottom left quadrant; and (3)
  PAH-dominated Seyferts (some non-HBLR Seyfert 2s and Seyfert 1.8/1.9s) lie
  mostly in the right quadrants. Positive values of
  $\alpha_{\lambda}(20\textrm{--}30)$ indicate red and steep long-wavelength
  continua and a stronger contribution from cold ($T \sim
  50\textrm{--}80\kel$) thermal dust components (see the bottom panels of
  Figure~1). This cold component is likely associated with star formation
  \citep{2001A&A...379..823K,2006ApJ...649...79S}. In the discussions that
  follow, we propose that the mid-IR spectra contain contribution from at
  least three thermal components: (1) a high temperature
  ($T\sim$~500--1200$\kel$) ``hot'' component dominating at shorter
  wavelengths, (2) a ``warm'' component ($T\sim$~200$\kel$) dominating at
  15--20$\mum$, and (3) a ``cool'' component ($T\sim$~50--80$\kel$) associated
  with star formation as mentioned above.
  
  The arrows in Figure~\ref{fig:continua} represent how the spectral shape
  changes from one quadrant to another. Positive values of
  $\alpha_{\lambda}(5.5\textrm{--}14.7)$ suggest a dominance of warm ($T\sim
  200\kel$) thermal components and relatively steep short-wavelength continua
  in $F_{\lambda}$. Objects that will fall in this quadrant will show enhanced
  15--20$\mum$ bump in the mid-IR spectra. This emission bump is likely due to
  a warm thermal component peaking at these wavelengths. We would like to
  clarify that this is not the $18\mum$ silicate feature, but rather a thermal
  modified blackbody (dust) emission underneath it, the short-wavelength side
  of which is clearly visible only when the short-wavelength continuum is
  weak. No object with the 15--20$\mum$ bump as the dominant feature in the
  mid-IR spectrum shows $10\mum$ in emission; this suggests that the
  short-wavelength continuum due to the hot component and any associated
  silicate emission features is being absorbed.
  
  As the emission of the hot dust component increases, the spectra begin to
  show a power-law-like flattening at short wavelengths:
  $\alpha_{\lambda}(5.5\textrm{--}14.7)$ becomes more negative and the
  dominance of the warm thermal continuum decreases. This power law is likely
  the Rayleigh--Jeans tail of the hot component with dust close to its
  sublimation temperature. At any point along this type 2 to type 1 sequence,
  adding a cold dust component to the spectrum shifts the spectrum toward
  positive (steep) $\alpha_{\lambda}(20\textrm{--}30)$. NGC~7469, a Seyfert
  1.5 with strong starburst contribution is identified in
  Figure~\ref{fig:continua}. Beyond this point in the right quadrants, we see
  that there are no Seyfert 1s and the region with
  $\alpha_{\lambda}(20\textrm{--}30) \gtrsim -0.5$ is mostly populated by
  Seyfert 1.8/1.9s and Seyfert 2s with undetected polarized broad emission
  lines. In these galaxies, the AGN continuum contribution is weaker than the
  contribution from the circum-nuclear starburst in the host galaxy
  \citep{2007ApJ...671..124D}. The spectra are primarily dominated by PAH
  features and very red long-wavelength continua. Mrk~938 and NGC~3079 show
  very red long-wavelength continua in our sample, and both of these galaxies
  are highly inclined to our line of sight (see Table~\ref{tab:objdata}),
  their continua are dominated by the emission from the cold dust. The
  relatively steep short-wavelength continua of these starburst-dominated
  Seyferts indicate that continuum measurements at $5.5\mum$ include a
  contribution from nearby $6.2\mum$ PAH complex.
  
  Figure~\ref{fig:opt-depth} compares the apparent silicate optical depth at
  $9.7\mum$ ($\tau_{9.7}$) with $\alpha_{\lambda}(5.5\textrm{--}14.7)$. We
  find a weak correlation between $\tau_{9.7}$ and
  $\alpha_{\lambda}(5.5\textrm{--}14.7)$. For $\tau_{9.7} \leq 0.4$, we find a
  Spearman rank correlation, $\rho = 0.46$ with $P_{null} = 1.85 \times
  10^{-5}$. This suggests that as the inclination of observer's line of sight
  changes from pole-on for type 1 objects to edge-on for type 2 objects, the
  inner hot dust is obscured, in agreement with the unified models
  \citep{1993ARA&A..31..473A}. This result is in agreement with trends noted
  by \citet{2007ApJ...655L..77H} between continuum colors and silicate
  strength.

  Measurement of optical depth requires defining a continuum which is a
  subjective process, leading to large errors in measured values of optical
  depths. We generate the continuum by fitting a spline curve through certain
  pivot points. As mentioned above, pivots at 5.5, 14.7, 20 and 30$\mum$ are
  used. Apart from these pivots, we also use additional points around 8 and
  12$\mum$ to define the blue and red ends of the silicate absorption feature.
  The process is very similar to the one presented in Figure~2 of
  \citet{2007ApJ...654L..49S} and should produce identical results in most
  cases. As a test, in Figure~\ref{fig:tau_pah_test}, we present a comparison
  between the 11.3$\mum$ PAH equivalent width and the optical depth. The
  absence of any correlation in the figure suggests that we are not biasing
  our continuum placements when strong PAH bands are present in the spectra.
  Thus, our optical depth measurements should be fairly accurate in most
  cases.
  
  A few objects in Figure~\ref{fig:opt-depth} do not follow this trend and
  show strong silicate optical depths $\tau_{9.7} > 0.5$. In all such objects,
  the strong silicate absorption is a result of absorption due to dust in the
  host galaxy rather than in the immediate vicinity of the central source.
  This conclusion is supported by objects that show very red long-wavelength
  continua (NGC~3079 and Mrk~938). To investigate the dependence of
  $\tau_{9.7}$ on the inclination of the host galaxy, we plotted $\tau_{9.7}$
  against the $b/a$ of the host galaxy. This comparison is shown in
  Figure~\ref{fig:opt-depth-ba}, which indicates that for $b/a < 0.5$, dust in
  the host galaxy disk can contribute significantly to the observed silicate
  absorption and the long-wavelength continuum. A few high-inclination
  galaxies (\eg IC4329A) do not show silicate absorption, but rather silicate
  emission. We checked {\it Hubble Space Telescope} images of IC4329A and note
  that the narrow-line region of this object is clearly visible in the F533N
  (narrow-band optical filter around [O~{\footnotesize III}] $\lambda5007$)
  filter above the inclined host galaxy disk. Our line of sight to the central
  source in IC4329A likely does not intersect any dense dust clouds in the
  host interstellar medium. An alternative explanation, when our line of sight
  is not close to the polar axis of the ionization bicone in an inclined host
  galaxy disk, could be that clumpy distributions can produce weak silicate
  emission features, or very weak absorption by filling in the existing
  absorption. In summary, the host interstellar medium can have substantial
  effect on the mid-IR nuclear continuum and a clear indication of this is a
  very red long-wavelength continuum in the mid-IR spectrum of an inclined
  disk galaxy. A comparison of $\alpha_{\lambda}(5.5\textrm{--}14.7)$ with
  $\tau_{9.7}$ for objects with $b/a > 0.5$ preserves the correlation in
  Figure~\ref{fig:opt-depth}, suggesting that the observed correlation is
  probing dust obscuration closer to the central source than on kilo--parsec
  scales.
  
  The effects of the host galaxy ISM on AGN mid-IR spectra are demonstrated in
  Figure~\ref{fig:ex-spectra}, where we plot representative mid-IR spectra
  around the silicate optical depth ($\tau_{9.7}$) versus
  $\alpha_{\lambda}(5.5\textrm{--}14.7\mum)$ plot from
  Figure~\ref{fig:opt-depth}. On the left-hand side, as the
  $\alpha_{\lambda}(5.5\textrm{--}14.7)$ increases from $-2.0$ to $1.0$, the
  short wavelength continuum is gradually suppressed, and we progress from
  strong type 1 source (bottom left) dominating the short-wavelength mid-IR to
  strong type 2 source (top left) peaking between 15 and 20$\mum$. The
  left-hand side panels show the true behavior of mid-IR active nuclear
  continua without contribution from external host galaxy features. As the
  apparent optical depth increases above 0.5, the spectra show stronger
  contribution from dust in the host galaxy, and enhanced PAH features (top
  right and bottom right in Figure~\ref{fig:opt-depth}). NGC~1194 (left,
  middle box in Figure~\ref{fig:ex-spectra}) lacks these starburst features,
  but shows one of the strongest silicate absorption features at both 10 and
  18$\mum$. Figures~\ref{fig:opt-depth-ba} and \ref{fig:ex-spectra} validate
  the observation that many Seyfert galaxies with highly inclined host galaxy
  disks are classified as type 2 or type 1.8/1.9s
  \citep{1980AJ.....85..198K,1995ApJ...454...95M}.
  
  These results imply that to study effects of dust in the torus, we need to
  look at sources with $-0.3 \lesssim \tau_{9.7} \lesssim 0.3$ and
  $\alpha_{\lambda}(20\textrm{--}30) \lesssim -0.5$. A key property that
  distinguishes active nuclear continua is the similarity of
  $\alpha_{\lambda}(20\textrm{--}30)$ for both type 1 and type 2 Seyferts (see
  Figure~\ref{fig:all_spec}), and the fact that the 20--30$\mum$ continua are
  bluer than the star formation dominated very red continua. For type 1.5 and
  type 2 Seyferts (S1h), the dominant 15--20$\mum$ feature suggests that the
  dust structures responsible for these features are warm ($T \sim 200\kel$),
  and the spectral index in the 20--30$\mum$ is decided by the slope of the
  Rayleigh--Jean's tail of this warm component. As can be seen from the top
  left panels in Figure~\ref{fig:ex-spectra}, the short-wavelength continua
  are significantly weaker in type 2 sources as compared to type 1 sources
  (bottom left). In the 20--30$\mum$ region of the spectrum, the differences
  are smaller in agreement with previous observation by
  \citet{2006AJ....132..401B}.
    
  \subsection{Non-HBLR Seyferts and Starburst Contributions}
  
  Based on the spectro-polarimetry survey of the CfA and the 12$\mum$ sample
  of Seyfert galaxies, \citet{2001ApJ...554L..19T,2003ApJ...583..632T}
  suggested that there are two different populations of Seyfert galaxies: (1)
  type 2 Seyferts that host a HBLR (S1h in Table~\ref{tab:objdata}), and hence
  are intrinsically identical to a type 1 Seyfert; and (2) type 2 Seyferts
  that show no signatures of the BLR in spectro-polarimetry (non-HBLR) and
  have optical narrow-line ratios weaker than the HBLRs on the BPT diagram
  \citep{1987ApJS...63..295V,1981PASP...93....5B}. Hence, they are closer to
  the sequence of star-forming galaxies on the BPT diagram. The term
  ``non-HBLR'' does not imply a lack of the BLR, but that it is not detected
  in the observations that lead to their classification as S2. For example,
  Mrk~573 was originally classified as a non-HBLR. It has been shown to
  contain a HBLR by \citet{2004AJ....128..109N} with higher signal-to-noise
  observations. Further work by \citet{2004MNRAS.348.1451L} and
  \citet{2007A&A...473..369H} suggests that non-HBLR objects are weak AGN with
  a dominant host galaxy component. Their mid-IR spectra should then be
  dominated by star-forming features within the host galaxies, which is what
  we find here.
  
  Five out of the 16 HBLR Seyferts that we added to our original sample show
  the enhanced emission bump around 15--20$\mum$ and the rest show power-law
  like mid-IR continua similar to most Seyfert 1s. One out of 16 shows strong
  PAH bands, and all of them have moderately strong $10\mum$ silicate
  absorption features. Thus, our comparison here provides qualitative
  agreement with the picture of non-HBLRs as weak AGN with a strong starburst
  component. Since we have complete mid-IR spectra, we can quantify starburst
  and AGN contributions over the whole IRS spectral range.

  \section{Starburst Contribution in Seyfert Galaxies}
  
  Many studies have explored the question of separating starburst contribution
  from AGN contribution using different diagnostics.
  \citet{1998ApJ...498..579G} compared the strength of $7.7\mum$ PAH to
  [O~{\footnotesize IV}]~$25.89\mum$/[Ne~{\footnotesize II}]~$12.81\mum$
  emission line ratio, to study the AGN contribution in ultra-luminous
  infrared galaxies (ULIRGs). \citet{2000A&A...359..887L} used spectral
  templates of H{\footnotesize II} regions, photo-dissociation regions and AGN
  to construct diagnostic diagrams. \citet{2002A&A...393..821S} suggested use
  of emission lines and continuum diagnostics. \citet{2008ApJ...689...95M}
  used estimate of [Ne~{\footnotesize II}] from pure AGN sources to constrain
  AGN and starburst fraction in a sample of Seyfert galaxies.
  \citet{2008MNRAS.385L.130N} used templates for starburst and AGN components
  in the 5--8$\mum$ region to estimate relative strengths in ULIRGs.
  
  In \citet{2007ApJ...671..124D}, we noted that the equivalent width of
  $6.2\mum$ PAH increases as $\alpha_{\lambda}(20\textrm{--}30\mum)$ became
  more positive, which suggested that this relation was driven by the
  starburst content of the active galaxy. In Figure~\ref{fig:continua}, we
  find that most Seyfert 1.8/1.9 and Seyfert 2 galaxies dominated by PAH
  emission have positive $\alpha_{\lambda}(20\textrm{--}30\mum)$. To measure
  the contribution of starburst features to individual galaxies on this
  diagram, we use the average starburst galaxy spectrum from
  \citet{2006ApJ...653.1129B}. We assume that the PAH inter-band ratios in
  this average spectrum are representative of typical PAH inter-band ratios in
  starburst galaxies. We are aware that PAH inter-band ratios can be different
  within different star-forming template spectra
  \citep[\eg][]{2004ApJ...613..986P,2007ApJ...656..770S}, and depend
  particularly on the strength of the starburst ionization field. Further,
  \citet{2006ApJ...653L..13S} also show that PAH ratios can be different in
  LINER spectra. We find similar variations in our Seyfert spectra also. We
  note that the strength of the 6.2, 7.7, and the 8.6$\mum$ PAH complexes vary
  much more than others. As a first-order comparison, assuming a fixed PAH
  template is acceptable. We also assume that the contribution of cool dust to
  the long-wavelength continuum in this average spectrum is typical of
  star-forming galaxies. This simple subtraction of a scaled starburst
  template assumes that any obscuration in the object is entirely due to the
  torus and not due to the surrounding starburst. We normalize the template
  starburst spectrum, so that the peak flux density of the 6.22 PAH band is
  unity. Then, we scale and subtract this template by trial and error from the
  Seyfert spectra. We require that the $17\mum$ PAH complex and the $11.3\mum$
  PAH bands be cleanly subtracted. Figure~\ref{fig:sb-sub} shows examples of
  such subtractions.
    
  Out of 107 sources for which we performed the starburst subtraction, 50 show
  power-law-like continuum over the whole IRS range with weak silicate
  emission features, after the subtraction. Twenty four objects show silicate
  absorption at $10\mum$. Seven objects show strong silicate emission features
  with $f_{10\mum} > f_{18\mum}$ and an underlying power-law-like continuum.
  We find 26 objects where the 15--20$\mum$ emission is much more prominent
  than the $10\mum$ emission.
  
  The majority of Seyfert galaxies in this sample show similar PAH inter-band
  ratios as in the starburst template spectrum. However, in a few cases like
  Mrk 477 in Figure~\ref{fig:sb-sub}, we find that the PAH $6.2$ and $7.7\mum$
  bands are over-subtracted. This indicates that PAH inter-band ratios in AGN
  spectra are not always similar to those in starburst galaxies. Assuming that
  the $11.3\mum$ and $17\mum$ bands represent the actual strength of the
  starburst in these sources, we find a deficit of emission at $6.2$ and
  $7.7\mum$ PAH bands in some Seyfert galaxies. On the other hand, in sources
  clearly dominated by starburst contribution such as Mrk~938 and NGC~3079, we
  find an excess of emission at $6.2$ and $7.7\mum$, after PAH bands at $11.3$
  and $17\mum$ are cleanly subtracted. We note that the over-subtraction of
  PAH $6.2$ and $7.7\mum$ bands occurs in Seyfert galaxies with weak
  short-wavelength mid-IR continua. These variations in PAH ratios are
  interesting, as relating the change in PAH inter-band ratios with the
  decreasing intensity of the interstellar radiation field over time, as the
  starburst fades, will provide crucial constraints on the time required for
  the active nucleus to become recognizable after the starburst episode.
  
  The starburst subtraction process essentially yields the AGN spectrum devoid
  of any star-forming features. These starburst-subtracted spectra are shown
  in Figure~\ref{fig:all_spec_nosb}. It is instructive to compare this figure
  with Figure~\ref{fig:all_spec}. Note the striking similarity of mid-IR
  active nuclear continua between all Seyfert optical subtypes. Type 1 spectra
  are flatter at short wavelengths than type 2 and type 1.8/1.9 spectra.
  Emission lines are generally weaker in PAH-dominated Seyfert 2s and Seyfert
  1.8/1.9s (see Figure~\ref{fig:all_spec}) suggesting weaker active nuclear
  continua \citep{2007ApJ...671..124D,2008ApJ...682...94M}.

  We measure the contribution from the starburst subtracted AGN spectrum at
  continuum wavelengths of 5.5, 8, 10, 14.7, 20, and 30$\mum$. In
  Figure~\ref{fig:lum1}, we plot the luminosity density at 5.5 and 20$\mum$
  from the active nucleus and the starburst component. Starburst contributions
  at 10, 1, 0.1, 0.01, and 0.001 times the active nucleus contribution are
  also shown with thin lines. The starburst-to-active nuclear continuum ratios
  are given in Table~\ref{tab:sbratio} for reference. The galaxies with strong
  PAH bands (almost all Seyfert 1.8/1.9s and some Seyfert 2s with undetected
  polarized broad lines) indeed have starburst luminosities almost as much as
  their active nucleus contribution, but not larger at this wavelength. The
  sample as a whole is weighted toward significant contribution ($\sim$
  factor of 10) from the active nucleus component at 5.5$\mum$.
  Figure~\ref{fig:lum1} (right) also shows similar comparison at 20 $\mum$.
  This figure shows that the AGN contribution decreases rapidly at longer
  wavelengths. Around $\sim20\mum$, the starburst and the active nuclear
  contribution are similar. This validates our use of
  $\alpha_{\lambda}(20\textrm{--}30\mum)$ in Figure~\ref{fig:continua} to
  separate objects dominated by starburst contribution.
  
  We construct spectral indices, as previously described in Section~3, from
  these starburst-subtracted spectra. In Section~3, we could not use the
  $8\mum$ flux density to construct $\alpha_{\lambda}(5.5\textrm{-}8.0\mum)$
  due to the contribution of the $7.7$ and $8.6\mum$ PAH bands. With the
  starburst-subtracted spectra this is now possible. A comparison of
  $\alpha_{\lambda}(5.5\textrm{-}8.0\mum)$ with
  $\alpha_{\lambda}(20\textrm{-}30\mum)$ shows that the long-wavelength
  continua are now flatter after subtraction of the starburst component (see
  Figure~\ref{fig:after-sb-sub}, top). We find that $\langle
  \alpha_{\lambda}(5.5\textrm{-}8\mum) \rangle = -0.73\pm0.07$ and $\langle
  \alpha_{\lambda}(20\textrm{-}30\mum) \rangle = -2.05\pm0.10$. For
  comparison, the average values for these quantities in
  Figure~\ref{fig:continua} are $-0.66\pm0.06$ and $-1.04\pm0.10$,
  respectively. There is clearly a large change in
  $\alpha_{\lambda}(20\textrm{-}30\mum)$, confirming our conclusion again from
  last paragraph. The main effect of subtracting the starburst template
  spectrum is to shift the PAH-dominated Seyfert 1.8/1.9s and Seyfert 2s to
  the left. Note that there are no Seyferts with
  $\alpha_{\lambda}(20\textrm{-}30\mum) > -0.5$ in the bottom right quadrant
  now, which proves our earlier point (see Section~3) about PAH contamination
  at $5.5\mum$. Figure~\ref{fig:after-sb-sub} (top) shows that there are at
  least two types of dust distributions in the active nuclear region, one that
  generates the short-wavelength continua and other that generates the
  long-wavelength continua. The major difference between these two dust
  distributions is likely to be their mean temperatures. By subtracting the
  starburst and associated cool dust contribution, we have essentially removed
  the starburst component that contributes most to the variety of Seyfert
  spectra \citep{2006AJ....132..401B}.

  A comparison of $\alpha_{\lambda}(5.5\textrm{-}8\mum)$ against luminosity
  density at $5.5\mum$ (Figure~\ref{fig:after-sb-sub}, bottom) shows that
  Seyfert 1s have $\langle \alpha_{\lambda}(5.5\textrm{-}8\mum) \rangle =
  -1.13 \pm 0.08$, while Seyfert 2s (type S1h, S1.8, S1.9 and S2) have
  $\langle \alpha_{\lambda}(5.5\textrm{-}8\mum) \rangle = -0.49 \pm 0.07$. A
  Kolmogorov--Smirnov test of the two distributions gives a probability of
  $4.42 \times 10^{-5}$ of null hypothesis that the two samples are drawn from
  the same distribution. The value of $\alpha_{\lambda}(5.5\textrm{-}8\mum)$
  for Seyfert 1s is in agreement with estimates derived for type 1 quasars
  \citep{2007ApJ...661...30G,2003A&A...402...87H}.

  \section{Summary}
\begin{figure}[!ht]
    \begin{center}
      \includegraphics[width=\textwidth]{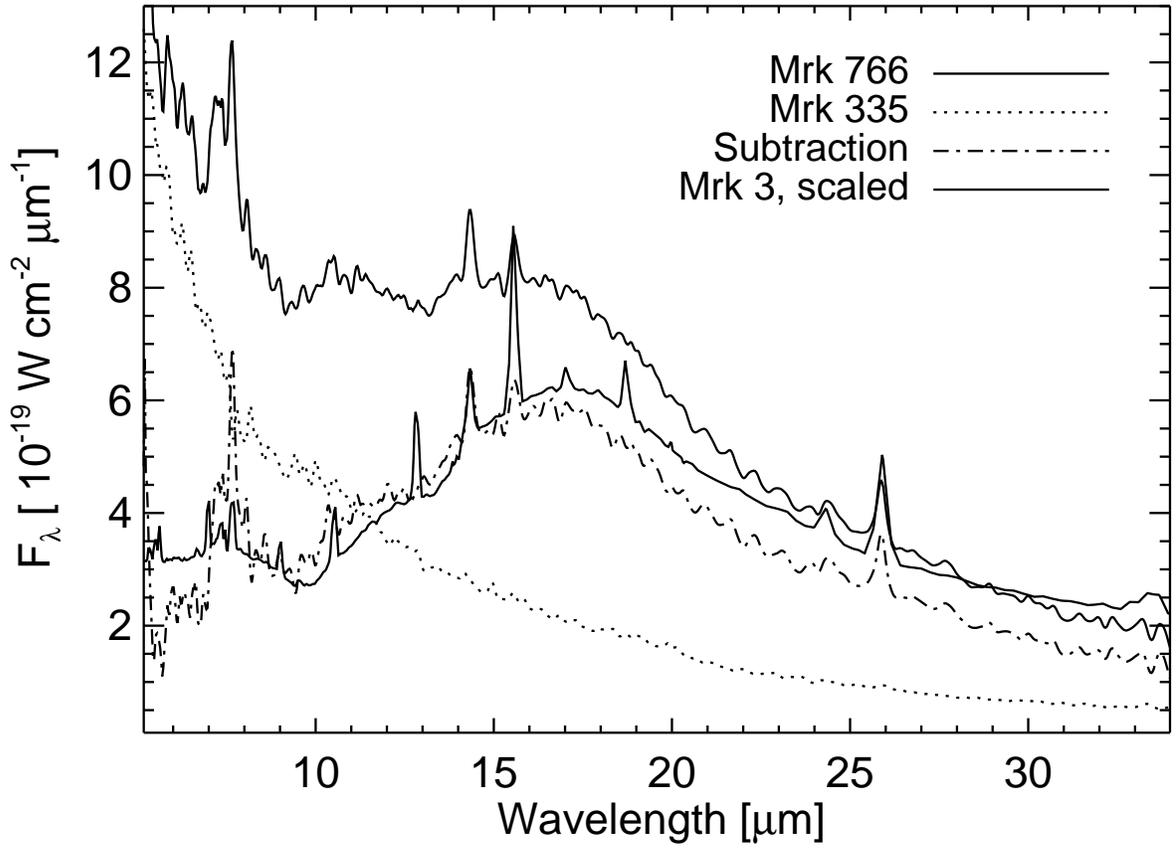}
    \end{center}
    \caption{Subtraction of Mrk 335 (type 1) spectrum from the spectrum of Mrk
      766 (type 1, starburst-subtracted, see Section~5). The residual spectrum
      is very similar to the Mrk 3 (type 2) spectrum and shows the prominent
      15--20$\mum$ bump. The spectra are displayed in the same order at
      10$\mum$ as is indicated in the legend.}
    \label{fig:bump}
  \end{figure}

\begin{figure}[!ht]
    \begin{center}
      \includegraphics[width=\textwidth]{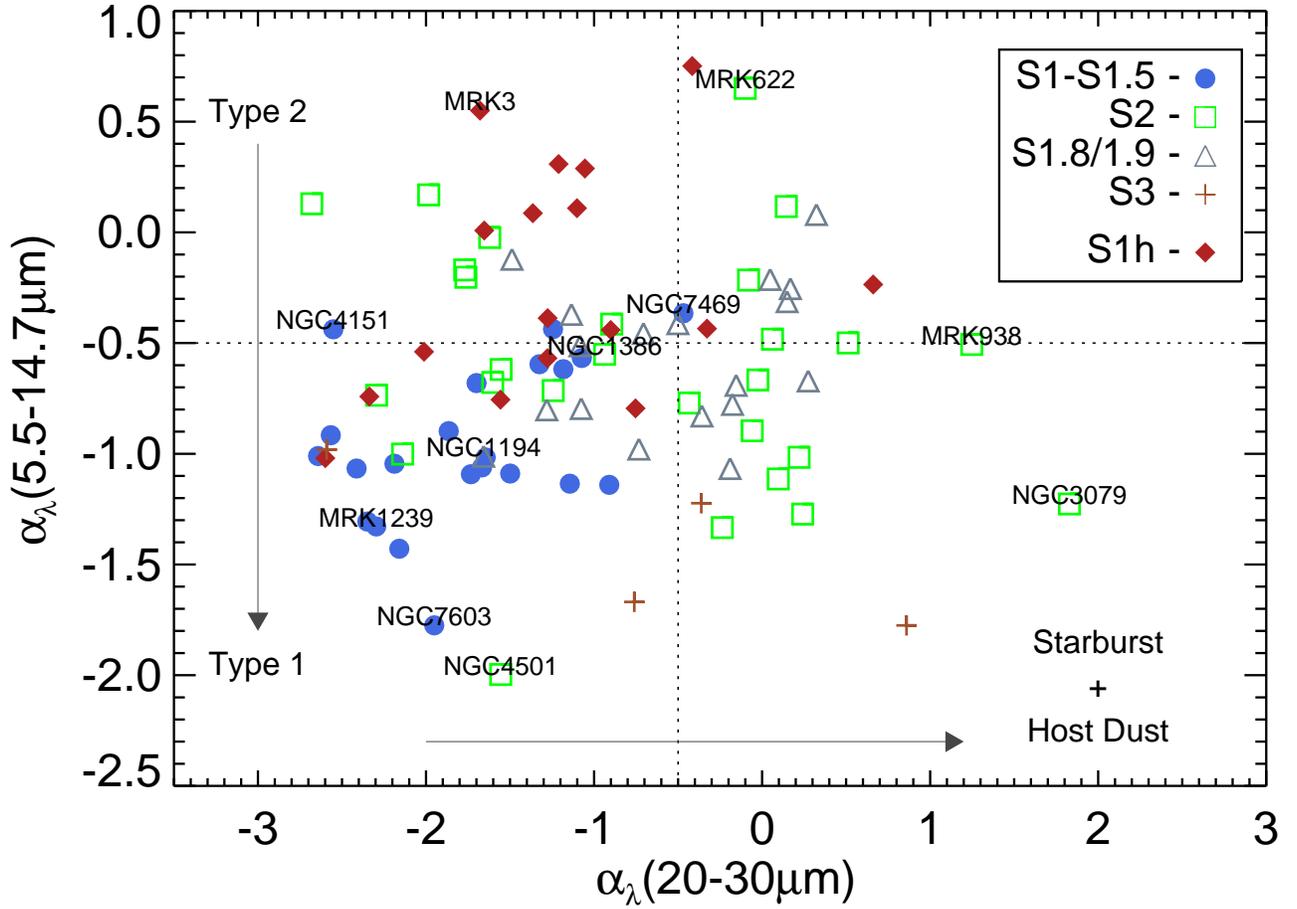}
    \end{center}
    \caption{Mid-IR continua of Seyfert Galaxies: Seyfert 2s with hidden
      broad-line regions (S1h) are represented as filled diamond symbols,
      Seyfert 1s (including 1.2s and 1.5s) are filled circles, Seyfert
      1.8/1.9s are open triangles, and Seyfert 2s with undetected polarized
      broad emission lines are open squares. Liners are represented with cross
      symbols. As $\alpha_{\lambda}(20\textrm{--}30\mum)$ increases, spectra
      contain more starburst contribution; as
      $\alpha_{\lambda}(5.5\textrm{--}14.7\mum)$ varies from 0.5 to -2.0,
      spectra are more dominated by the hot dust component due to the active
      nucleus. The arrows indicate how the spectral shape changes from type 2
      to type 1 Seyferts (see also Figure~\ref{fig:ex-spectra}); and along
      this sequence, addition of a cold starburst component of increasing
      strength moves the source position to the right in the figure. The
      dotted lines show rough division between different Seyfert types. See
      Section~4 for further discussion.}
    \label{fig:continua}
  \end{figure}

\begin{figure}[!ht]
    \begin{center}
      \includegraphics[width=\textwidth,angle=0]{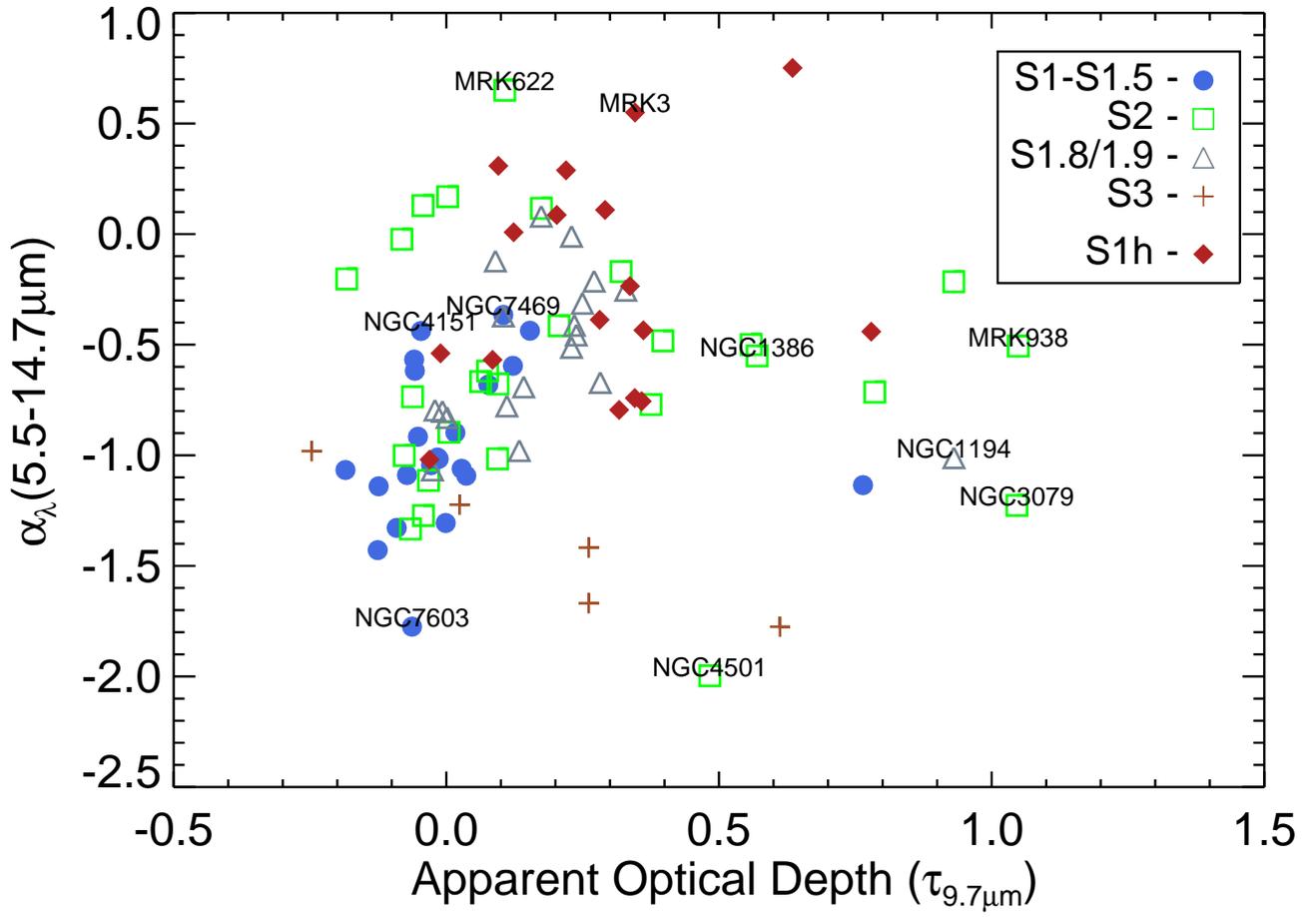}
    \end{center}
    \caption{There is a weak correlation between apparent optical depth
      ($\tau_{9.7}$) and $\alpha_{\lambda}(5.5\textrm{--}14.7\mum)$ providing
      general support for inclination dependence in AGN models. Some sources
      show large apparent silicate optical depth ($\tau_{9.7} \sim 1$): these
      have highly inclined host galaxy disks (see Table~\ref{tab:objdata}).
      For sources with $\tau_{9.7}\gtrsim0.4$ silicate absorption is primarily
      due to cold dust in the host galaxy.}
    \label{fig:opt-depth}
  \end{figure}

\begin{figure}[!ht]
    \begin{center}
      \includegraphics[width=\textwidth]{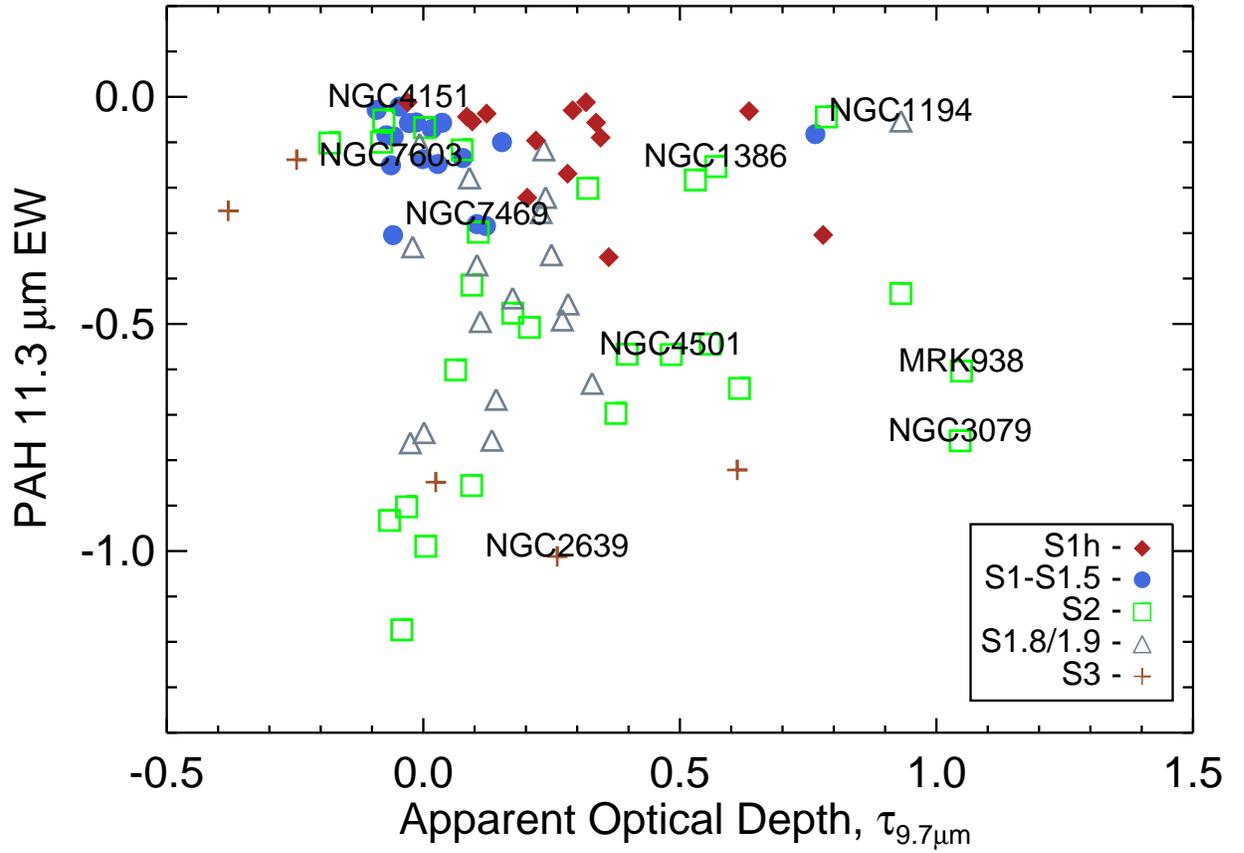}
    \end{center}
    \caption{There is no correlation between the equivalent width of the
      11.3$\mum$ PAH band and the measured apparent optical depth at
      9.7$\mum$, suggesting that the presence of strong PAH bands does not
      bias our continuum placement in the short wavelength region of the
      mid-IR spectra, where defining the continuum is highly subjective due to
      the presence of blended PAH emission bands, and silicate and ice
      absorption features.}
    \label{fig:tau_pah_test}
  \end{figure}

\begin{figure}[!ht]
    \begin{center}
      \includegraphics[width=\textwidth]{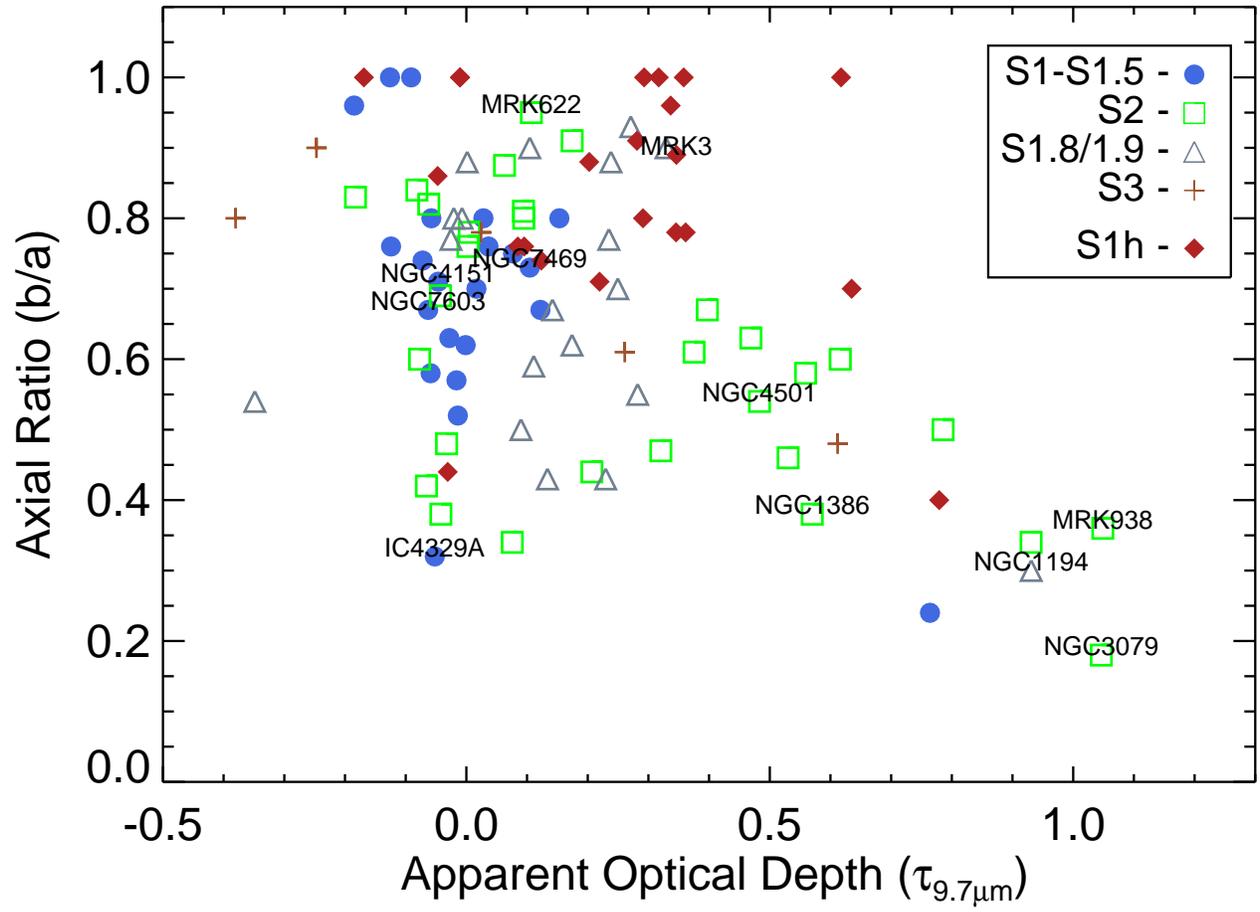}
    \end{center}
    \caption{Correlation between the optical depth of $9.7\mum$ silicate
      feature and the ratio of minor-to-major axis ($b/a$) of the host galaxy.
      The $b/a$ values are taken from the NASA/IPAC Extragalactic Database.
      Objects for which $b/a$ could not be estimated are assumed to have $b/a
      = 1$.}
    \label{fig:opt-depth-ba}
  \end{figure}

\begin{figure}[!ht]
    \begin{center}
      \includegraphics[width=\textwidth]{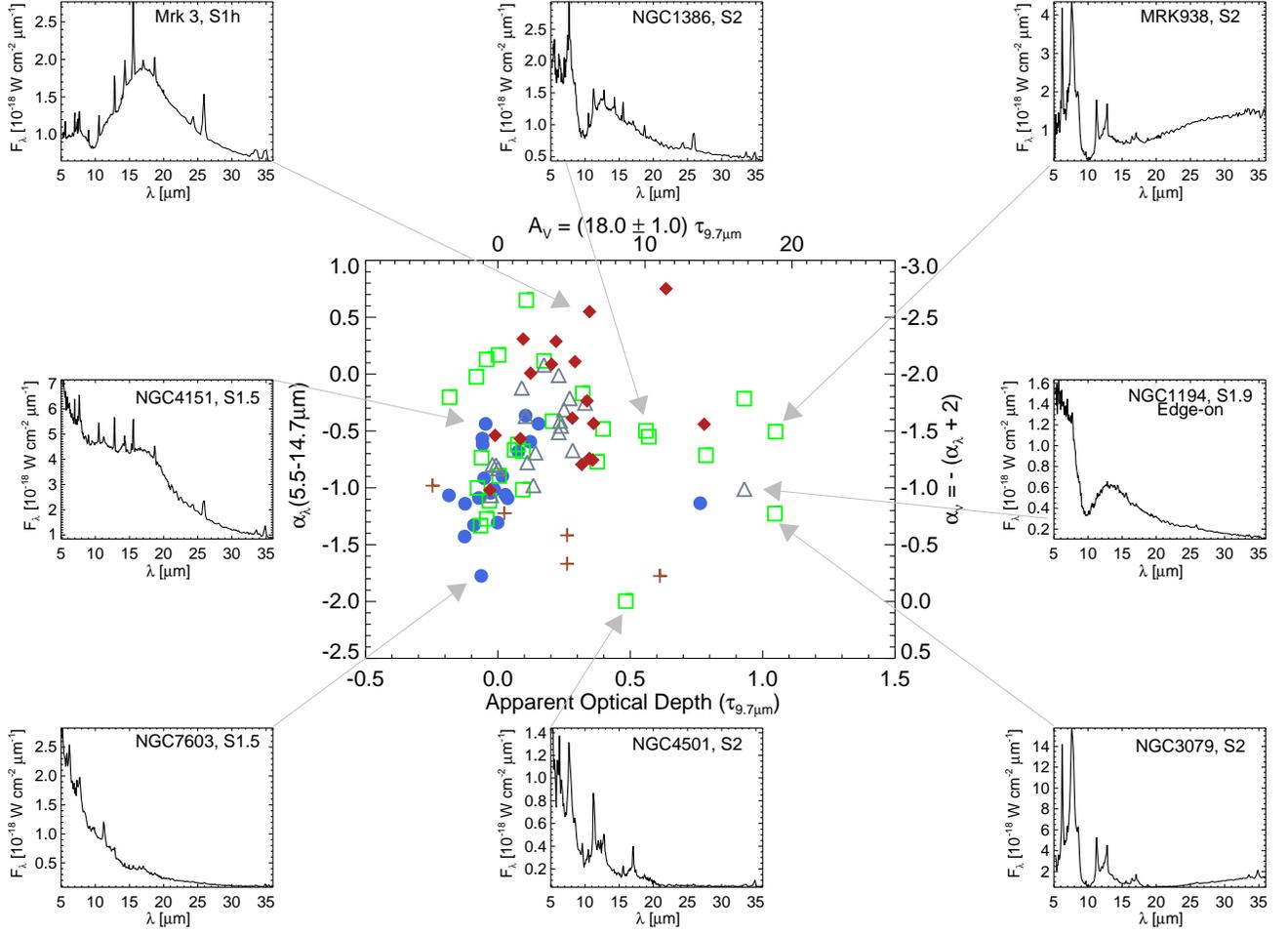}
    \end{center}
    \caption{Effect of host galaxy dust on mid-IR spectra of AGN---central
      plot shows the variation of $\alpha_{\lambda}(5.5\textrm{--}14.7\mum)$
      with $\tau_{9.7\mum}$; left panels: spectra of AGN-dominated galaxies;
      middle panels: increasing cold dust and PAH contribution attributed to
      star formation in the host galaxy; right panels: Inclined host galaxies
      without strong PAH features (NGC 1194) or with strong PAH features (Mrk
      938 and NGC 3079). Essentially, in all systems with $\tau_{9.7\mum}
      \gtrsim 0.4$ the apparent optical depth at $9.7\mum$ is likely due to
      cold dust in the host galaxy. In the panels, from left to right, star
      formation contribution increases.}
    \label{fig:ex-spectra}
  \end{figure}

\begin{figure}[!ht]
    \begin{center}
      \includegraphics[width=\textwidth]{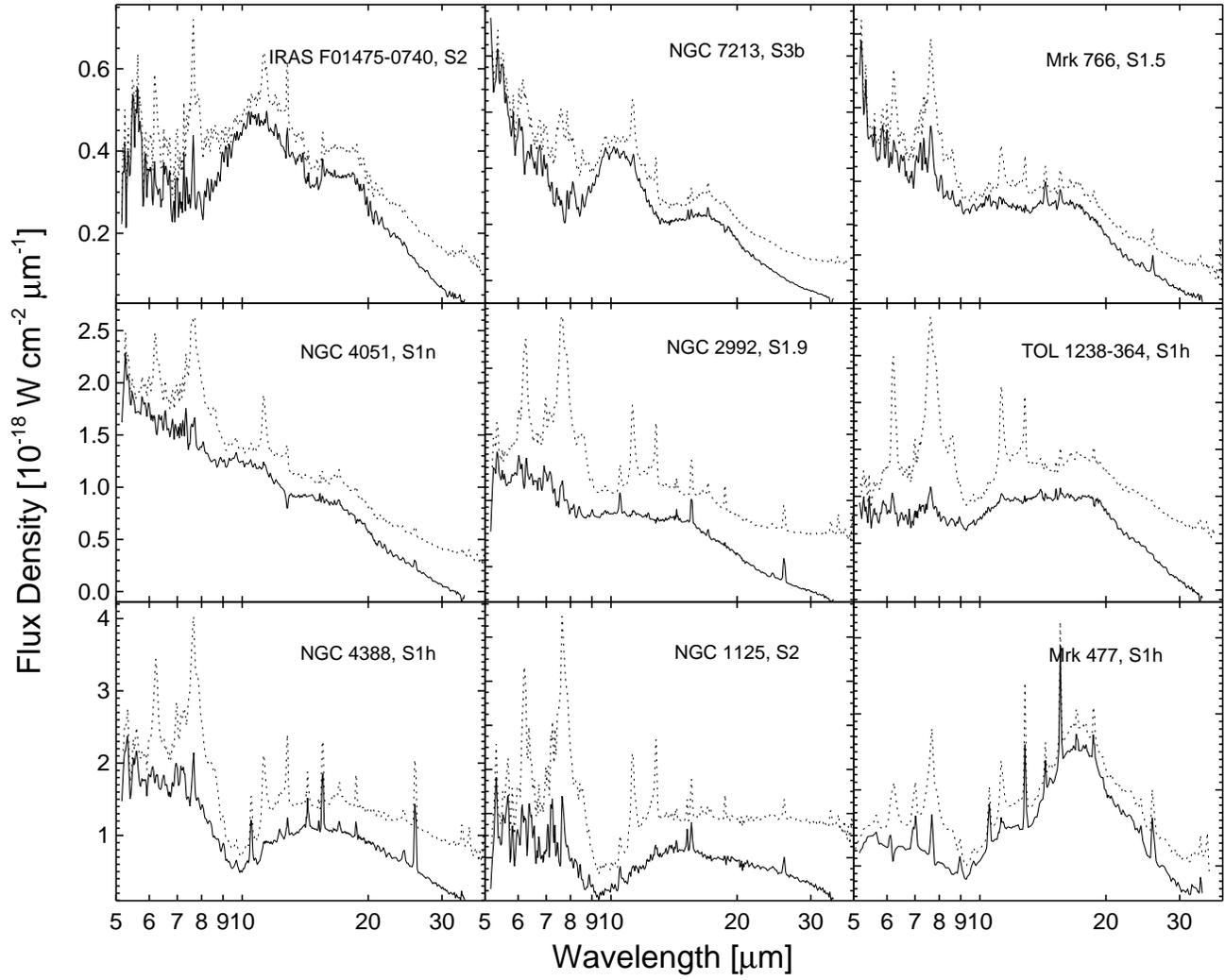}
    \end{center}
    \caption{Seyfert spectra after subtraction of starburst components: the
      dotted line shows the observed spectrum and the solid line shows the
      subtraction of an average starburst template from this spectrum. See
      Figure~1 in \citet{2007ApJ...671..124D} for identification of emission
      lines and PAH bands in mid-IR spectra. Note the clearly visible silicate
      features, the power-law-like shape of the 5--8$\mum$ continuum and the
      15--20$\mum$ bump. In most cases, the $7.65\mum$ [Ne~\textsc{VI}]
      lines are also cleanly separated.}
    \label{fig:sb-sub}
  \end{figure}

\begin{figure}[!ht]
    \begin{center}
      \includegraphics[width=\textwidth]{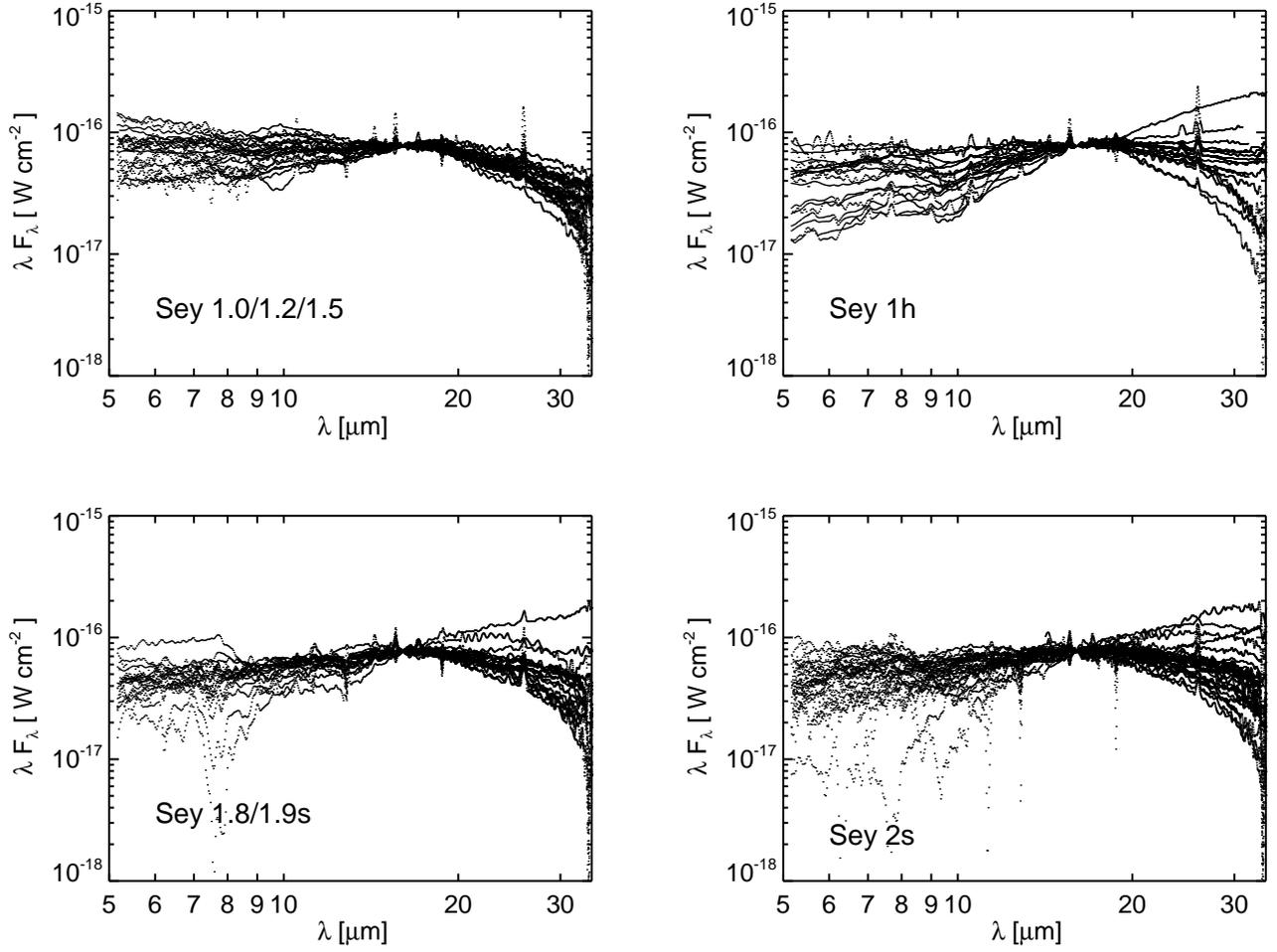}
    \end{center}
    \caption{Mid-infrared spectra from Figure~\ref{fig:all_spec} after
      subtraction of a scaled starburst template spectrum. The spectra are
      normalized at 14.7$\mum$. Top left: Seyfert 1s; top right: Seyfert 2s
      with evidence of broad optical emission lines in polarized light;
      bottom left: Seyfert 1.8/1.9s; bottom right: Seyfert 2s with
      undetected polarized broad emission lines. Compare this figure with
      Figure~\ref{fig:all_spec}. On average, after starburst subtraction,
      PAH-dominated Seyfert 1.8/1.9s and Seyfert 2s show similar continuum
      shapes as type 2 Seyferts with HBLR (top right panel). There is a
      striking similarity in the active nuclear continuum of all Seyfert types
      beyond $\sim$15$\mum$.}
    \label{fig:all_spec_nosb}
  \end{figure}

\begin{figure}[!ht]
    \begin{center}
      \includegraphics[width=0.5\textwidth]{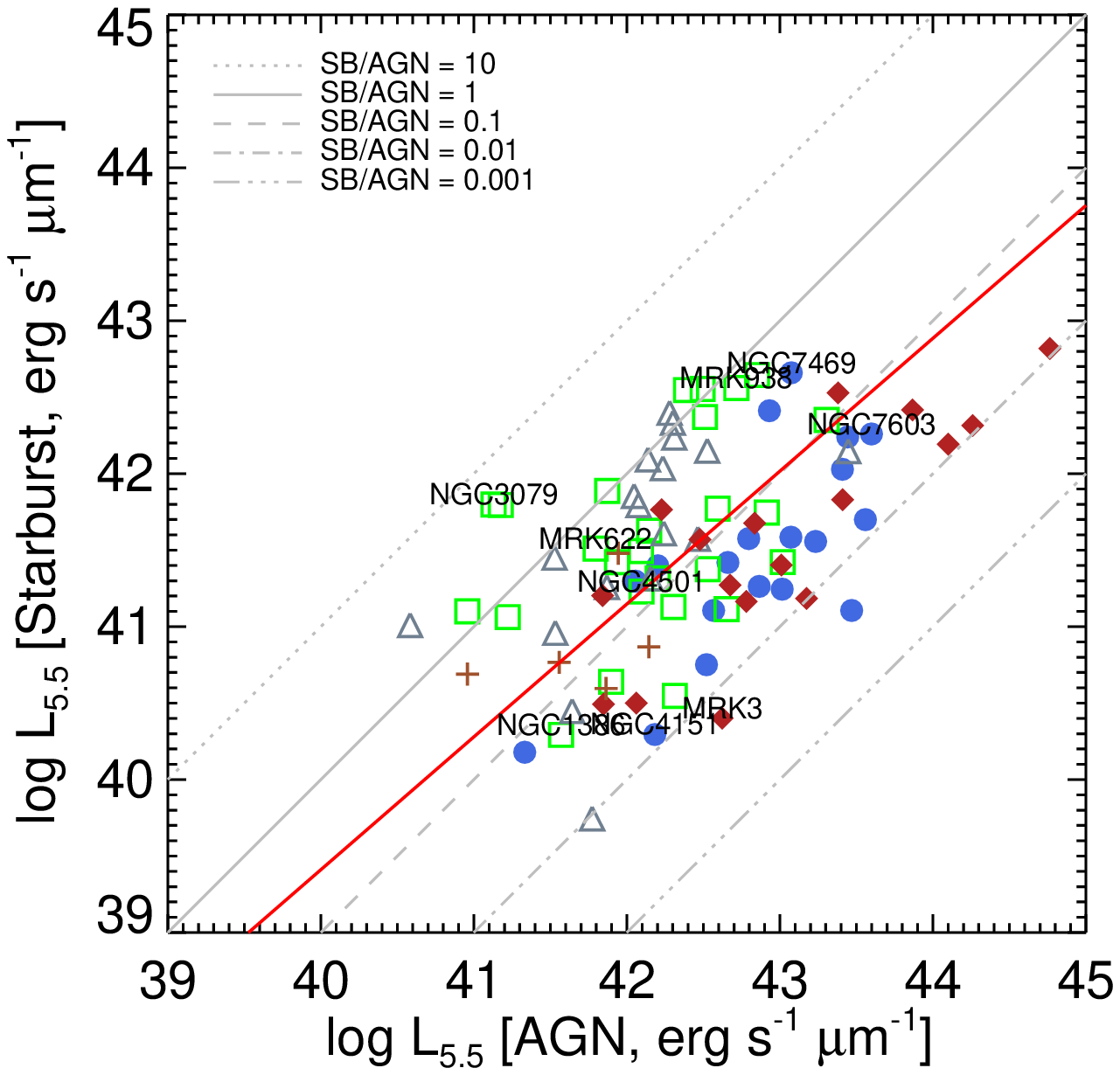}%
      \includegraphics[width=0.5\textwidth]{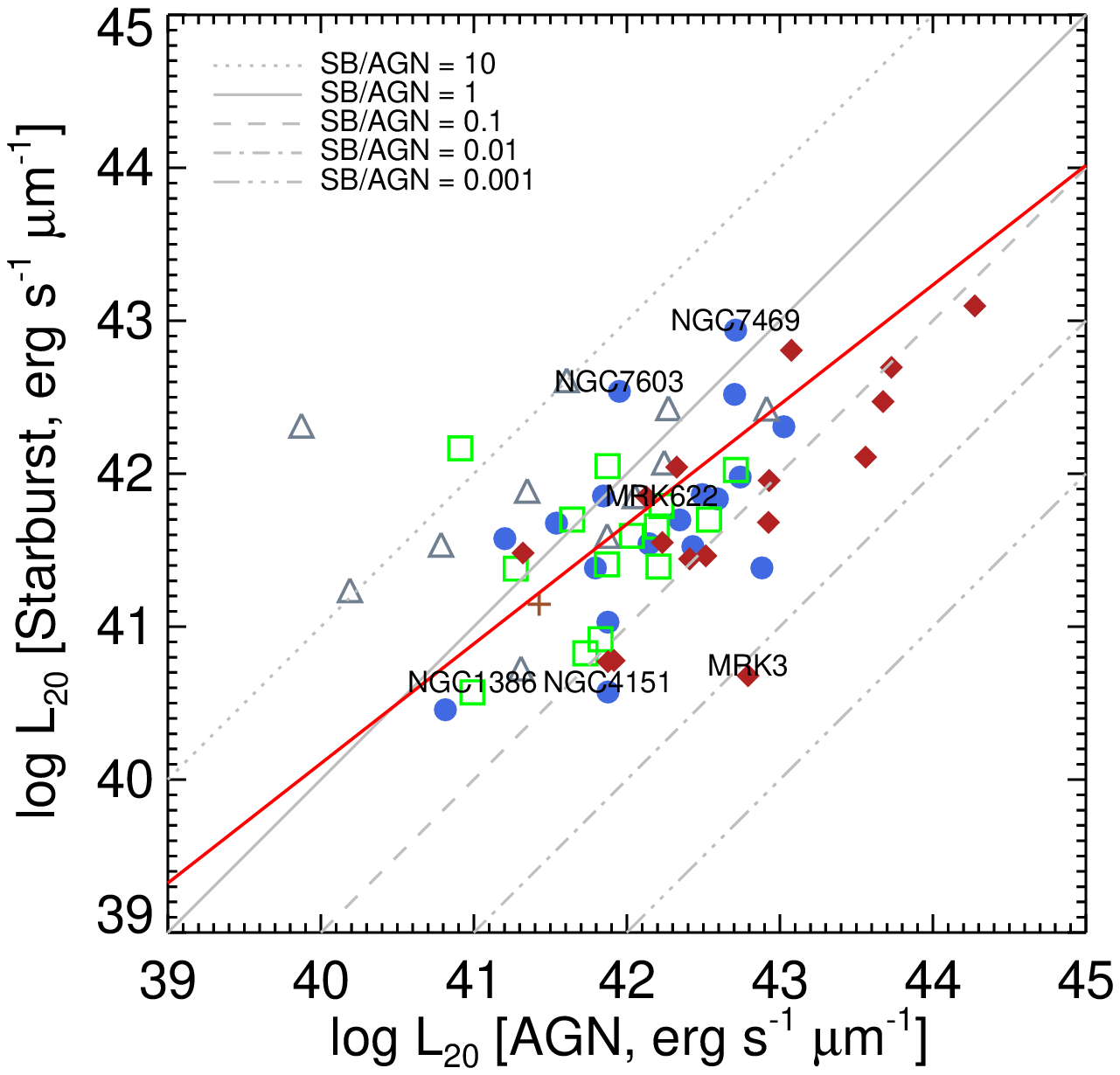}
    \end{center}
    \caption{Luminosity density at $5.5\mum$ (left) and 20$\mum$ (right) from
      the starburst and the active nuclear components. Our sample shows a
      strong contribution from the active nuclear component, and galaxies with
      large starburst contributions are preferentially Seyfert 1.8/1.9s and
      some Seyfert 2s with unidentified broad polarized emission lines. The
      thick red line shows a linear regression fit (bisector method) to data
      points. The diagonal lines show starburst-to-AGN contribution ratio.
      Symbols for Seyfert types are same as in Figure~\ref{fig:continua}. At
      longer wavelengths, the average starburst fraction increases and is
      roughly the same as the AGN contribution at $\sim 20\mum$.}
    \label{fig:lum1}
  \end{figure}

\begin{figure}[!ht]
    \begin{center}
      \includegraphics[width=0.8\textwidth]{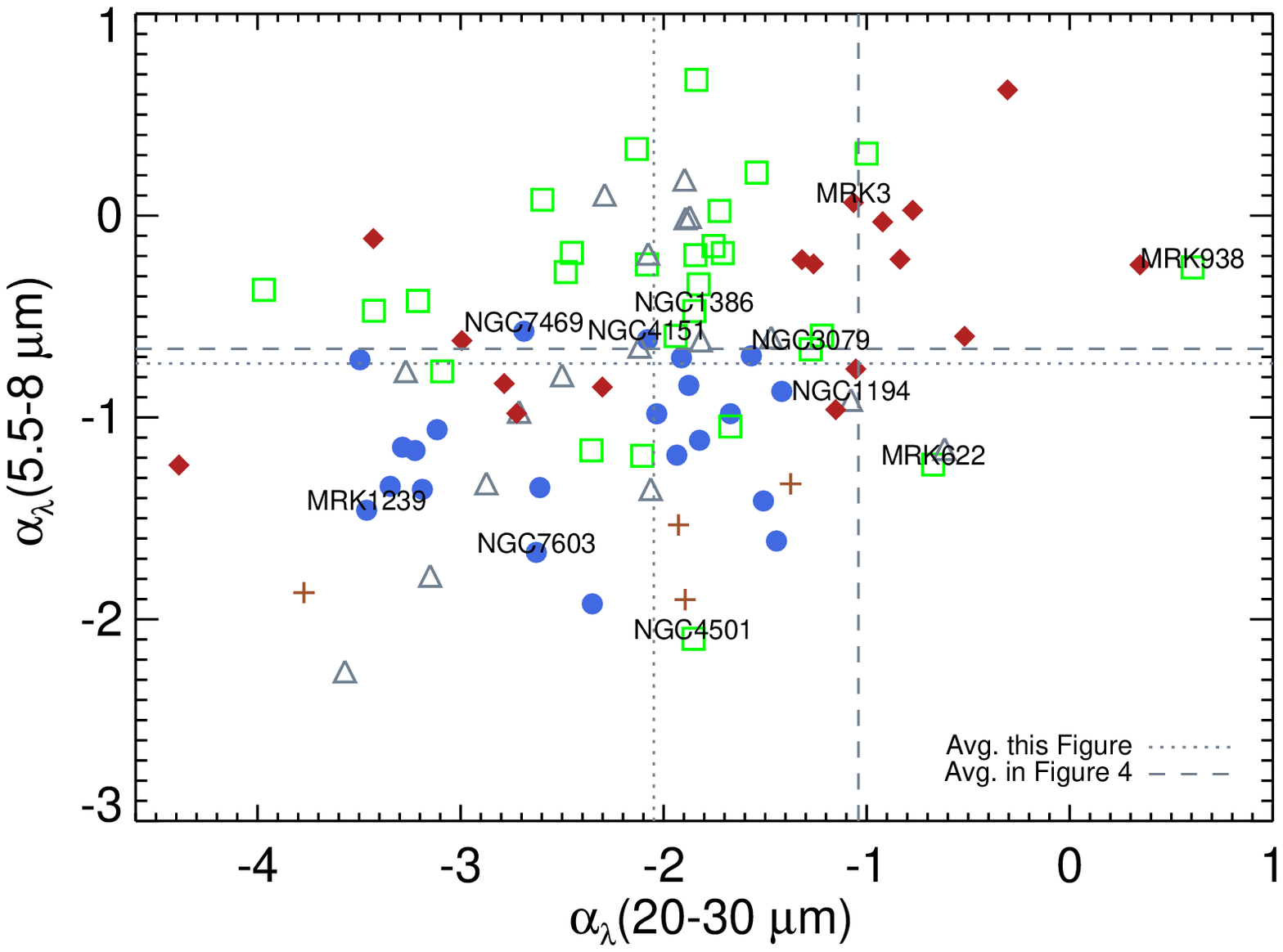}
      \includegraphics[width=0.8\textwidth]{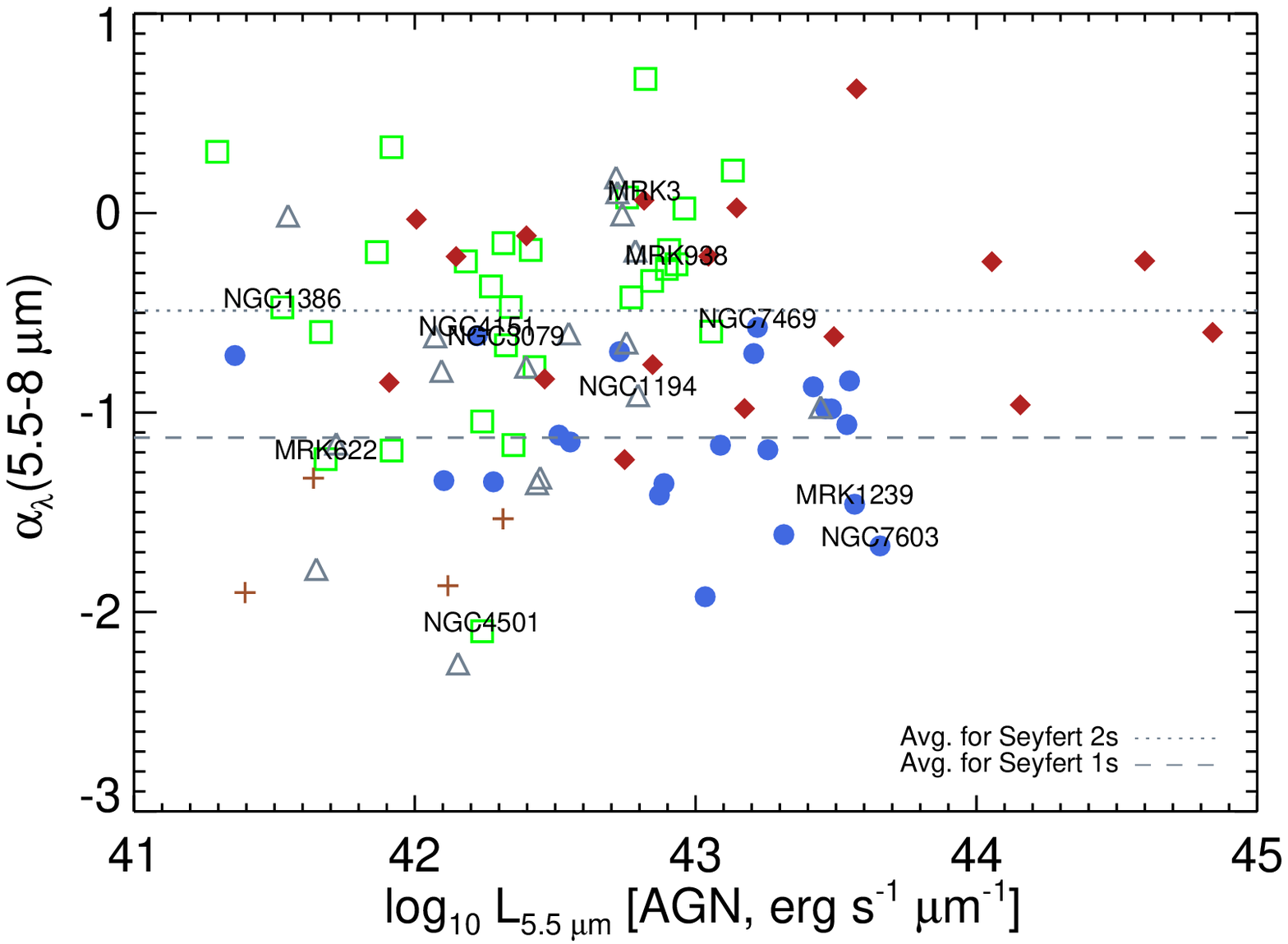}
    \end{center}
    \caption{Top: Comparison between $\alpha_{\lambda}(5.5\textrm{--}8.0\mum)$
      and $\alpha_{\lambda}(20\textrm{--}30\mum)$ after subtraction of
      starburst component. $\alpha_{\lambda}(20\textrm{--}30\mum)$ values are
      more negative than in Figure~\ref{fig:continua} (dashed lines),
      indicating flatter continua than those at short wavelength. Bottom:
      wavelength spectral index between 5.5 and $8\mum$ after starburst
      subtraction compared to the luminosity density of the AGN component at
      $5.5\mum$. The dashed line shows the average spectral index for Seyfert
      1 galaxies, and the dot-dashed line shows the average spectral index for
      Seyfert 2 (including S1h) and Seyfert 1.8/1.9 galaxies. Symbols for
      Seyfert types are same as in Figure~\ref{fig:continua}.}
    \label{fig:after-sb-sub}
  \end{figure}

  An analysis of archival {\it Spitzer Space Telescope} mid-IR spectra
  of Seyfert galaxies is presented. We focus on understanding the intrinsic
  shape of the active nuclear continuum in the mid-IR region and how it
  relates to other properties of the source such as the 10$\mum$ silicate
  optical depth. We assumed a template spectrum for the starburst component,
  and subtracted it from the Seyfert spectra to separate the active nuclear
  contribution from the circum-nuclear starburst contribution. Our primary
  conclusions from this study are as follows:
  \begin{enumerate}
  \item Seyfert spectra are classified effectively between AGN- and
    starburst-dominated categories based on the spectral indices,
    $\alpha_{\lambda}(5.5\textrm{--}14.7\mum)$ and
    $\alpha_{\lambda}(20\textrm{--}30\mum)$ (see Figure~\ref{fig:continua}).
    Seyferts dominated by the AGN contribution have flatter spectra with
    $\alpha_{\lambda}(5.5\textrm{--}14.7\mum) \sim -1.13$. The added starburst
    contribution from the host galaxy in the large {\it Spitzer} aperture (see
    Table~\ref{tab:objdata}) makes $\alpha_{\lambda}(20.0\textrm{--}30.0\mum)$
    more positive or steeper. A key property that distinguishes AGN-dominated
    spectra is that the 20--30$\mum$ continuum is flatter
    ($\alpha_{\lambda}\sim -2$) than the very steep 20--30$\mum$ continuum of
    starburst-dominated objects ($\alpha_{\lambda}\sim 1$). The 20--30$\mum$
    continuum is likely formed from the Rayleigh--Jeans tail of the ``warm''
    dust component. It is likely that there are multiple ``warm'' components
    with different temperatures. Further, Type 2 Seyferts with polarized broad
    emission lines in their optical spectra (type S1h) show
    $\alpha_{\lambda}(5.5\textrm{--}14.7\mum) \sim - 0.49$ much different than
    average type 1 Seyferts that show
    $\alpha_{\lambda}(5.5\textrm{--}14.7\mum) \sim -1.13$. Note the steeper
    and weak short-wavelength continuum in Figure~\ref{fig:all_spec_nosb}, as
    compared to type 1 Seyferts. This is a direct evidence for presence of the
    dust torus that blocks our view of the hot dust closer in.
  \item After starburst subtraction, Seyfert 1.8/1.9s and Seyfert 2s with
    strong PAH features in their spectra show similar active nuclear continuum
    as Seyfert 2s with weak/absent PAH features in their mid-IR spectra and
    polarized broad emission lines in their optical spectra (see
    Figure~\ref{fig:all_spec_nosb}). This suggests presence of similar
    quantities and/or properties of dusty material around the central
    accretion disk in these type 2 sources. \citet{2003ApJ...583..632T} had
    proposed existence of two types of Seyfert 2s: the HBLR and the non-HBLRs.
    We compared spectral indices in the 5--8$\mum$ region after starburst
    subtraction and find that both the HBLR and non-HBLR show similar spectral
    indices (Figure~\ref{fig:after-sb-sub}, bottom), suggesting similarity
    rather than differences between the two classifications. While, non-HBLRs
    tend to show stronger starburst contribution as compared to their AGN
    contribution, the converse (that starburst-dominated systems lack BLR
    signatures) is not necessarily true. The additional host galaxy
    contribution likely complicates the identification of BLR in these
    systems. As we show in Figure~\ref{fig:continua}, simple continuum indices
    effectively separate AGN-dominated Seyferts from starburst-dominated
    Seyferts in the mid-IR.
  \item \citet{2007ApJ...671..124D} showed that a large part of the silicate
    absorption in some Seyfert galaxies comes from starburst-heated cold dust
    in the host galaxy rather than the dust torus. This conclusion was based
    on the fact that only highly inclined galaxies showed large silicate
    optical depths. Here, we put that result on a better statistical basis by
    presenting a correlation between the $b/a$ and the measured optical depth
    for this sample of 109 sources (see Figure~\ref{fig:opt-depth-ba}). All
    objects with significant optical depth are highly inclined and are also
    classified as Seyfert 2s or Seyfert 1.8/1.9s. This confirms previous
    results by \cite{1980AJ.....85..198K} and \cite{1995ApJ...454...95M}, and
    highlights the importance of considering the host galaxy contribution in
    concealing AGN.
  \item On average, Seyfert galaxies dominated by the AGN continuum tend to
    show weak silicate absorption ($\tau_{9.7} \lesssim 0.4$). The short
    wavelength continuum index ($\alpha_{\lambda}(5.5\textrm{--}14.7\mum)$)
    and the apparent silicate optical depth $\tau_{9.7}$
    (Figure~\ref{fig:opt-depth}) may to be correlated in AGN-dominated
    objects. Seyfert optical types form a continuous sequence of increasing
    optical depth along this correlation from type 1s, type 1.8/1.9s, to type
    2s with HBLRs. This validates the general inclination dependence inherent
    in AGN models noted before in the mid-IR by \citet{2007ApJ...655L..77H}.
    But, as can be noted in Figure~\ref{fig:opt-depth}, there is not a strict
    relationship between the strength of silicate features and the optical
    spectral type.
  \item Figure~\ref{fig:after-sb-sub} (top) shows that there are at least two
    types of dust distributions in the active nuclear region, one that
    generates the short-wavelength continua and another that generates the
    long-wavelength continua. The major difference between these two dust
    distributions is likely to be their mean temperatures. By subtracting the
    starburst and associated cool dust contribution, we have essentially
    removed the starburst component that contributes most to the variety in
    Seyfert spectra \citep{2006AJ....132..401B}. In Figure~\ref{fig:bump}, we
    demonstrate the existence of this ``warm'' component which dominates the
    long-wavelength continuum, by separating it from the hot dust component in
    Mrk 766. This simple template subtraction exercise provides proof that
    Seyfert spectra are primarily thermal in nature and composed of at least
    three thermal components with $T\sim$ 1000, 200, and 60$\kel$. The
    reported ``break'' in the spectrum at $\sim$ 20$\mum$ in type 1 Seyfert
    spectra is a result of this warm component being brighter at $\sim 20
    \mum$ than the Rayleigh--Jeans tail of the hot component. The
    above-mentioned similarity of continua beyond $\sim15\mum$ in
    AGN-dominated sources is also due to this warm dust component being
    present in almost all observed AGN spectra.
  \end{enumerate}
  
  \acknowledgments

  We would like to thank Nadia Zakamska for insightful comments on an early
  draft of this paper. G.T.R. acknowledges support from an Alfred P. Sloan
  Research Fellowship. This work is based on archival data obtained with the
  {\it Spitzer Space Telescope}, which is operated by the Jet Propulsion
  Laboratory, California Institute of Technology under a contract with NASA.
  This research has made use of the NASA/IPAC Extragalactic Database (NED)
  which is operated by the Jet Propulsion Laboratory, California Institute of
  Technology, under contract with the National Aeronautics and Space
  Administration. This research has also made use of NASA's Astrophysics Data
  System Bibliographic Services. The IRS was a collaborative venture between
  Cornell University and Ball Aerospace Corporation funded by NASA through the
  Jet Propulsion Laboratory and Ames Research Center. SMART was developed at
  Cornell University and is available through the Spitzer Science Center at
  Caltech.

  {\it Facility:} \facility{{\it Spitzer}}

  %
  \clearpage
  \LongTables
  \begin{deluxetable}{lclclcrrcr}
    \tabletypesize{\tiny}
    \tablewidth{0pt}
    \tablecolumns{10}
    \tablecaption{Observation Summary, Redshifts, Optical Seyfert Types and Extraction Apertures.}
    \tablehead{\colhead{Galaxy} &
      \colhead{Redshift} &
      \colhead{Seyfert} &
      \colhead{$b/a$} &
      \colhead{{\it Spitzer}} &
      \colhead{Obs.} &
      \multicolumn{3}{c}{Extraction Rectangle} &
      \colhead{Parsec} \\
      \colhead{Name} &
      \colhead{} &
      \colhead{Type} &
      \colhead{} &
      \colhead{AOR} &
      \colhead{Mode} &
      \colhead{R.A. (deg.)} &
      \colhead{Dec. (deg.)} &
      \colhead{Aperture ($\arcsec$)} &
      \colhead{per $\arcsec$} \\
      \colhead{(1)} &
      \colhead{(2)} &
      \colhead{(3)} &
      \colhead{(4)} &
      \colhead{(5)} &
      \colhead{(6)} &
      \colhead{(7)} &
      \colhead{(8)} &
      \colhead{(9)} &
      \colhead{(10)}
    }
    \startdata
             3C 226  &  0.817700 &   S1h &   1.00 &  11298560  & S & $\cdots$    & $\cdots$    & $\cdots$         &  24655.19\\
             3C 234  &  0.184800 &   S1h &   1.00 &  11305728  & S & $\cdots$    & $\cdots$    & $\cdots$         &   4292.75\\
             3C 265  &  0.811000 &   S1h &   1.00 &  11299584  & S & $\cdots$    & $\cdots$    & $\cdots$         &  24406.85\\
             3C 321  &  0.096100 &   S1h &   0.70 &  10828544  & S & $\cdots$    & $\cdots$    & $\cdots$         &   2110.22\\
        CGCG381-051  &   0.03067 &    H2 &   0.67 &  12483328  & M &  357.174904 &    2.237392 &  18.506x27.508   &    643.03\\
         ESO 12-G21  &  0.030021 &  S1.5 &   0.22 &  12465920  & M &   10.178473 &  -79.235550 &  16.650x24.890   &    629.12\\
          ESO 33-G2  &   0.01810 &    S2 &   0.82 &  12473088  & M &   74.014906 &  -75.539219 &  16.651x24.891   &    375.94\\
          FSC 09104  &  0.442000 &   S1h &   1.00 &   6619136  & S & $\cdots$    & $\cdots$    & $\cdots$         &  11694.26\\
           IC 4329A  &   0.01605 &  S1.2 &   0.29 &  12480000  & M &  207.334737 &  -30.307954 &  12.957x18.319   &    332.84\\
            IC 5063  &  0.011348 &   S1h &   0.66 &  12465152  & M &  313.002802 &  -57.064322 &  25.401x35.919   &    234.49\\
    IRAS 05189-2524  &  0.042563 &   S1h &   0.96 &   4969216  & S & $\cdots$    & $\cdots$    & $\cdots$         &    900.21\\
   IRASF 01475-0740  &   0.01767 &    S2 &   0.83 &  12474880  & M &   27.512248 &   -7.431963 &  14.799x22.275   &    366.89\\
   IRASF 03450+0055  &  0.031000 &  S1.5 &   0.96 &  12455424  & M &   56.918450 &    1.084858 &  16.640x24.881   &    650.11\\
   IRASF 04385-0828  &   0.01510 &    S2 &   0.50 &  12447232  & M &   70.229938 &   -8.374696 &  16.657x23.555   &    312.92\\
   IRASF 15480-0344  &   0.03030 &    S2 &   0.84 &  12480256  & M &  237.676552 &   -3.886854 &  12.955x18.318   &    635.10\\
        MCG+0-29-23  &  0.024897 &    S2 &   0.88 &  12443648  & M &  170.300469 &   -2.984643 &  14.801x19.668   &    519.76\\
      MCG-03-34-064  &  0.016541 &   S1h &   0.80 &  20367616  & S & $\cdots$    & $\cdots$    & $\cdots$         &    343.15\\
        MCG-2-33-34  &   0.01463 &   S1n &   0.58 &  12481280  & M &  193.056402 &  -13.413451 &  12.952x18.317   &    303.07\\
         MCG-2-40-4  &   0.02519 &    S1 &   0.81 &  12481536  & M &  237.108125 &  -13.755182 &  14.804x22.282   &    525.99\\
         MCG-2-8-39  &   0.02989 &    S2 &   0.69 &  12476928  & M &   45.129079 &  -11.417929 &  14.791x23.687   &    626.32\\
        MCG-3-34-63  &   0.02133 &    S2 &   0.29 &  12477696  & M &  200.584610 &  -16.706397 &  16.651x22.278   &    444.10\\
         MCG-3-58-7  &   0.03146 &  S1.9 &   0.80 &  12457984  & M &  342.405623 &  -19.275421 &  14.799x20.927   &    659.98\\
        MCG-5-13-17  &   0.01264 &  S1.5 &   0.67 &  12468480  & M &   79.899175 &  -32.659368 &  12.947x18.308   &    261.45\\
        MCG-6-30-15  &   0.00775 &    S2 &   0.60 &  12457472  & M &  203.980171 &  -34.293642 &  16.646x23.544   &    159.71\\
           Mrk 1239  &  0.019927 &   S1n &   0.38 &  12453120  & M &  148.079113 &   -1.612924 &  16.649x23.546   &    414.45\\
              Mrk 3  &  0.013509 &   S1h &   0.89 &   3753472  & S & $\cdots$    & $\cdots$    & $\cdots$         &    279.61\\
            Mrk 334  &   0.02196 &  S1.8 &   0.70 &  10870016  & S & $\cdots$    & $\cdots$    & $\cdots$         &    457.44\\
            Mrk 335  &   0.02578 &  S1.2 &   1.00 &  12476416  & M &    1.582227 &   20.201022 &  16.662x23.556   &    538.55\\
            Mrk 348  &   0.01503 &   S1h &   0.78 &  12472832  & M &   12.197642 &   31.954601 &  16.652x24.891   &    311.45\\
           Mrk 463E  &  0.050000 &   S1h &   1.00 &   4980736  & S & $\cdots$    & $\cdots$    & $\cdots$         &   1063.21\\
            Mrk 471  &   0.03423 &  S1.8 &   0.67 &  10868992  & S & $\cdots$    & $\cdots$    & $\cdots$         &    719.56\\
            Mrk 477  &  0.037726 &   S1h &   0.71 &  17643008  & S & $\cdots$    & $\cdots$    & $\cdots$         &    795.09\\
              Mrk 6  &   0.01881 &  S1.5 &   0.63 &  12483584  & M &  103.064183 &   74.431101 &  16.653x23.552   &    390.89\\
            Mrk 609  &   0.03449 &  S1.8 &   0.90 &  10870528  & S & $\cdots$    & $\cdots$    & $\cdots$         &    725.17\\
            Mrk 622  &   0.02323 &    S2 &   0.95 &  10869248  & S & $\cdots$    & $\cdots$    & $\cdots$         &    484.35\\
            Mrk 704  &  0.029234 &  S1.2 &   0.57 &  12444416  & M &  139.607652 &   16.304607 &  18.497x26.162   &    612.27\\
            Mrk 766  &  0.012929 &  S1.5 &   0.80 &  12465408  & M &  184.608565 &   29.814439 &  16.650x23.549   &    267.49\\
             Mrk 79  &   0.02219 &  S1.2 &   0.52 &  12453632  & M &  115.635502 &   49.809536 &  16.651x24.887   &    462.31\\
            Mrk 817  &   0.03145 &  S1.5 &   0.80 &  12461056  & M &  219.101248 &   58.795290 &  18.501x24.890   &    659.77\\
            Mrk 883  &   0.03750 &  S1.9 &   0.62 &  10869504  & S & $\cdots$    & $\cdots$    & $\cdots$         &    790.20\\
              Mrk 9  &   0.03987 &  S1.5 &   0.80 &  12483072  & M &  114.235533 &   58.770872 &  16.649x23.548   &    841.60\\
            Mrk 938  &   0.01978 &    S2 &   0.36 &  12473856  & M &    2.778625 &  -12.109529 &  18.495x26.162   &    411.35\\
           NGC 1056  &   0.00515 &    S2 &   0.48 &  12464896  & M &   40.701906 &   28.571643 &  16.653x23.548   &    105.92\\
           NGC 1125  &   0.01093 &    S2 &   0.50 &  12454656  & M &   42.919692 &  -16.652879 &  18.501x26.167   &    225.78\\
         NGC 1143/4  &  0.028847 &    S2 &   0.63 &  12448512  & M &   43.801754 &   -0.185407 &  18.498x24.888   &    603.99\\
           NGC 1194  &   0.01360 &  S1.9 &   0.48 &  12472064  & M &   45.955575 &   -1.105563 &  14.802x22.279   &    281.51\\
           NGC 1241  &   0.01351 &    S2 &   0.61 &  12468224  & M &   47.812412 &   -8.924313 &  18.500x27.500   &    279.63\\
           NGC 1320  &   0.00888 &    S2 &   0.32 &  12454400  & M &   51.203760 &   -3.044172 &  16.659x23.551   &    183.15\\
           NGC 1365  &  0.005457 &  S1.8 &   0.55 &  12480768  & M &   53.402894 &  -36.143278 &  20.354x30.120   &    112.26\\
           NGC 1386  &  0.002895 &    S2 &   0.38 &  12474112  & M &   54.195896 &  -35.994983 &  16.653x22.280   &     59.44\\
           NGC 1667  &   0.01517 &    S2 &   0.78 &  12459520  & M &   72.155696 &   -6.323982 &  24.049x41.076   &    314.38\\
           NGC 2273  &  0.006138 &   S1h &   0.78 &   4851712  & S & $\cdots$    & $\cdots$    & $\cdots$         &    126.33\\
           NGC 2622  &   0.02862 &  S1.8 &   0.50 &  10869760  & S & $\cdots$    & $\cdots$    & $\cdots$         &    599.14\\
           NGC 2639  &   0.01113 &    S3 &   0.61 &  12477440  & M &  130.906163 &   50.203587 &  24.045x42.587   &    229.95\\
           NGC 2992  &  0.007710 &  S1.9 &   0.31 &  12478208  & M &  146.424140 &  -14.327517 &  20.342x27.495   &    158.88\\
           NGC 3079  &   0.00372 &    S2 &   0.18 &  12475136  & M &  150.502129 &   55.682721 &  20.352x31.506   &     76.42\\
           NGC 3081  &  0.007976 &   S1h &   0.76 &  18509824  & S & $\cdots$    & $\cdots$    & $\cdots$         &    164.39\\
           NGC 3227  &   0.00386 &  S1.5 &   0.67 &  12450304  & M &  155.875181 &   19.871115 &  20.323x36.634   &     79.31\\
           NGC 3511  &  0.003699 &    S1 &   0.34 &  12473600  & M &  165.848750 &  -23.089783 &  24.050x38.136   &     75.99\\
           NGC 3516  &   0.00884 &  S1.5 &   0.76 &  12473344  & M &  166.716783 &   72.571111 &  18.499x26.162   &    182.32\\
           NGC 3660  &  0.012285 &    S2 &   0.81 &  12477184  & M &  170.883777 &   -8.658924 &  16.647x23.541   &    254.04\\
           NGC 3786  &   0.00893 &  S1.8 &   0.59 &  10871808  & S & $\cdots$    & $\cdots$    & $\cdots$         &    184.19\\
           NGC 3982  &   0.00370 &  S1.9 &   0.88 &  12449536  & M &  179.130254 &   55.128503 &  24.054x32.734   &     76.01\\
           NGC 4051  &   0.00234 &   S1n &   0.75 &  12451072  & M &  180.797723 &   44.533906 &  18.503x27.506   &     48.02\\
           NGC 4151  &   0.00332 &  S1.5 &   0.71 &  12470784  & M &  182.642569 &   39.408036 &  16.648x24.887   &     68.18\\
            NGC 424  &   0.01166 &   S1h &   0.44 &  12444160  & M &   17.867777 &  -38.085092 &  20.347x28.774   &    241.00\\
           NGC 4388  &  0.008419 &   S1h &   0.19 &  12460288  & M &  186.443590 &   12.662449 &  20.347x28.778   &    173.58\\
           NGC 4501  &  0.007609 &    S2 &   0.54 &  12445440  & M &  187.994494 &   14.420342 &  20.351x30.112   &    156.78\\
           NGC 4507  &  0.011801 &   S1h &   0.76 &  18511104  & S & $\cdots$    & $\cdots$    & $\cdots$         &    243.94\\
           NGC 4579  &   0.00507 &   S3b &   0.80 &  12462080  & M &  189.439250 &   11.821549 &  24.051x36.721   &    104.26\\
           NGC 4593  &  0.009000 &  S1.0 &   0.74 &  12457216  & M &  189.913681 &   -5.344758 &  16.652x23.546   &    185.64\\
           NGC 4594  &  0.003416 &  S1.9 &   0.40 &  12456960  & M &  189.997069 &  -11.623687 &  16.648x23.545   &     70.16\\
           NGC 4602  &   0.00847 &  S1.9 &   0.35 &  12465664  & M &  190.161835 &   -5.127129 &  24.043x42.585   &    174.64\\
     NGC 4922 NED01  &  0.023860 &    S2 &   0.79 &  12477952  & M &  195.349342 &   29.317079 &  25.406x43.701   &    497.72\\
           NGC 4941  &   0.00370 &    S2 &   0.53 &  12471552  & M &  196.052465 &   -5.542940 &  25.400x47.925   &     76.01\\
           NGC 4968  &   0.00986 &    S2 &   0.47 &  12464128  & M &  196.781667 &  -23.674650 &  20.352x27.507   &    203.52\\
           NGC 5005  &   0.00316 &   S3b &   0.48 &  12475648  & M &  197.741665 &   37.061589 &  18.499x24.887   &     64.89\\
            NGC 513  &   0.01954 &  S1.9 &   0.43 &  12467712  & M &   21.112954 &   33.797022 &  20.350x28.780   &    406.29\\
           NGC 5135  &  0.013693 &    S2 &   0.67 &  12445696  & M &  201.439919 &  -29.831243 &  18.495x27.497   &    283.46\\
     NGC 5256 NED01  &  0.027600 &    S2 &   0.75 &  12459264  & M &  204.581404 &   48.278533 &  20.353x30.119   &    577.35\\
           NGC 526A  &   0.01910 &  S1.9 &   0.53 &  12454912  & M &   20.987415 &  -35.061446 &  25.401x47.921   &    397.01\\
           NGC 5347  &   0.00779 &    S2 &   0.76 &  12481792  & M &  208.329504 &   33.492829 &  14.798x22.273   &    160.54\\
           NGC 5506  &  0.006181 &   S1i &   0.24 &  12453888  & M &  213.317594 &   -3.205489 &  18.499x24.891   &    127.22\\
           NGC 5548  &   0.01717 &  S1.5 &   0.93 &  12481024  & S & $\cdots$    & $\cdots$    & $\cdots$         &    356.37\\
           NGC 5929  &   0.00831 &    S3 &   0.78 &  12444928  & M &  231.531375 &   41.672728 &  14.803x20.931   &    171.32\\
           NGC 5953  &   0.00656 &    S2 &   0.81 &  12476160  & M &  233.641348 &   15.196836 &  20.346x30.113   &    135.06\\
           NGC 6810  &  0.006775 &    S2 &   0.28 &  12479488  & M &  295.901492 &  -58.651454 &  18.498x26.161   &    139.51\\
           NGC 6860  &  0.014884 &  S1.5 &   0.62 &  12462592  & M &  302.204152 &  -61.095376 &  20.347x30.111   &    308.39\\
           NGC 6890  &  0.008069 &  S1.9 &   0.80 &  12452608  & M &  304.580525 &  -44.802776 &  18.497x27.499   &    166.32\\
           NGC 7130  &   0.01615 &  S1.9 &   0.93 &  12463616  & M &  327.082844 &  -34.953547 &  16.652x24.889   &    334.94\\
           NGC 7172  &   0.00868 &    S2 &   0.56 &  12450048  & M &  330.515750 &  -31.867704 &  25.400x39.675   &    179.00\\
           NGC 7213  &  0.005839 &   S3b &   0.90 &  12449024  & M &  332.323337 &  -47.161801 &  20.348x27.500   &    120.15\\
           NGC 7314  &   0.00476 &  S1.9 &   0.46 &  12469504  & M &  338.950394 &  -26.048111 &  25.397x39.677   &     97.87\\
           NGC 7469  &   0.01632 &  S1.5 &   0.73 &  12472320  & M &  345.815619 &    8.871944 &  16.648x22.273   &    338.51\\
           NGC 7496  &   0.00550 &    S2 &   0.91 &  12462336  & M &  347.449971 &  -43.430211 &  20.352x30.114   &    113.14\\
           NGC 7582  &  0.005254 &    S1 &   0.42 &  12445184  & M &  349.601583 &  -42.364982 &  20.349x28.777   &    108.06\\
           NGC 7590  &  0.005255 &    S2 &   0.37 &  12482560  & M &  349.731754 &  -42.233643 &  18.498x26.163   &    108.08\\
           NGC 7603  &   0.02952 &  S1.5 &   0.67 &  12450816  & M &  349.737435 &    0.241247 &  20.352x30.116   &    618.39\\
           NGC 7674  &   0.02892 &   S1h &   0.91 &  12468736  & M &  351.987312 &    8.776636 &  18.501x27.500   &    605.55\\
            NGC 788  &  0.013603 &   S1h &   0.74 &  18944512  & S & $\cdots$    & $\cdots$    & $\cdots$         &    281.57\\
            NGC 931  &   0.01665 &  S1.0 &   0.21 &  12460032  & M &   37.060954 &   31.309135 &  18.500x26.163   &    345.44\\
    SDSS J1039+6430  &  0.401776 &   S1h &   1.00 &  13630208  & S & $\cdots$    & $\cdots$    & $\cdots$         &  10444.00\\
    SDSS J1641+3858  &  0.595835 &   S1h &   0.86 &  13630464  & S & $\cdots$    & $\cdots$    & $\cdots$         &  16742.90\\
       TOL 1238-364  &   0.01092 &   S1h &   0.88 &  12466432  & M &  190.226517 &  -36.754040 &  16.654x23.549   &    225.58\\
    UGC 11680 NED01  &  0.025988 &    S2 &   0.80 &  12459008  & M &  316.924025 &    3.868699 &  18.504x26.167   &    542.97\\
          UGC 12138  &   0.02497 &  S1.8 &   0.88 &  10871296  & S & $\cdots$    & $\cdots$    & $\cdots$         &    521.31\\
           UGC 7064  &   0.02500 &  S1.9 &   0.90 &  12467456  & M &  181.186346 &   31.179312 &  16.661x23.556   &    521.95\\
             UM 146  &   0.01741 &  S1.9 &   0.77 &  10871040  & S & $\cdots$    & $\cdots$    & $\cdots$         &    361.42\\
 WIR-IRAS 23060+0505  &  0.173000 &   S1h &   1.00 &   4374528  & S & $\cdots$    & $\cdots$    & $\cdots$         &   3990.25\\
    \enddata    
    \tablecomments{Col. (1): Galaxy name; Col. (2): Redshift, obtained from
      NASA Extragalactic Database (NED); Col. (3): Seyfert type from
      \citet{2006A&A...455..773V}---S1: Seyfert 1 optical spectrum; S1h:
      broad polarized Balmer lines detected; S1i: broad Paschen lines observed
      in the infrared; S1n: narrow-line Seyfert 1; S1.0, S1.2, S1.5, S1.8, and
      S1.9:  intermediate Seyfert galaxies (Note: In this paper, we consider
      S1.0 to S1.5 to be Seyfert 1s in all analysis.); S2: Seyfert 2 spectrum;
      S3:  LINER; S3b:  LINER with broad Balmer lines; S3h: LINER with broad
      polarized Balmer lines detected; H2: nuclear H{\tiny II} region; Col.
      (4): Axial ratio, minor-to-major axis ratio of host galaxy obtained from
      NED, for host galaxies where $b/a$ was not available we assumed it to be
      1; Col. (5): {\em Spitzer} archive Astronomical Observation Request
      (AOR) number; Col. (6): Observing mode for {\em Spitzer/IRS} spectrum:
      S:  staring mode, M:  mapping mode; Col. (7 and 8):  Extraction
      rectangle for mapping-mode spectra: R.A. and Dec. of center-point in
      degrees; Col. (9): Extraction aperture in arc-seconds. Col. (10): Radial
      extent in parsecs per arc-second of the extraction aperture for the
      galaxy.}
    \label{tab:objdata}
  \end{deluxetable}
  \clearpage

  \LongTables
  \begin{deluxetable}{lcccccc}
    \tabletypesize{\tiny}
    \tablewidth{0pt}
    \tablecolumns{7}
    \tablecaption{Continuum Flux Densities and Apparent Optical Depth at $9.7\mum$.}
    \tablehead{\colhead{Galaxy Name} &
      \multicolumn{5}{c}{Continuum Flux Density (Jy)} &
      \colhead{Optical Depth} \\
      \colhead{} &
      \colhead{5.5$\mum$} &
      \colhead{10$\mum$} &
      \colhead{14.7$\mum$} &
      \colhead{20$\mum$} &
      \colhead{30$\mum$} &
      \colhead{9.7$\mum$} \\
      \colhead{(1)} &
      \colhead{(2)} &
      \colhead{(3)} &
      \colhead{(4)} &
      \colhead{(5)} &
      \colhead{(6)} &
      \colhead{(7)}
    }
    \startdata
    3C 226           &  3.82E-03   &  4.73E-03  &  1.57E-02   &  2.03E-02  &  $\cdots$  &   0.6175 \\
                     & (4.37E-04)  & (9.31E-04) & (1.02E-03)  & (1.98E-03) &            &  \\
    3C 234           &  5.25E-02   &  1.34E-01  &  2.21E-01   &  2.67E-01  &  2.66E-01  &  -0.0106 \\
                     & (4.35E-03)  & (6.46E-03) & (3.47E-03)  & (3.02E-03) & (6.53E-03) & \\    
    3C 265           &  6.66E-03   &  1.06E-02  &  1.61E-02   &  $\cdots$  &  $\cdots$  &  -0.1688 \\
                     & (1.50E-03)  & (1.19E-03) & (1.08E-03)  &            &            & \\
    3C 321           &  1.19E-02   &  3.28E-02  &  1.78E-01   &  2.71E-01  &  5.15E-01  &   0.6349 \\
                     & (1.42E-03)  & (5.12E-03) & (1.53E-02)  & (1.37E-02) & (2.16E-02) & \\
       CGCG381-051   &  8.73E-02   &  1.24E-01  &  1.93E-01   &  3.62E-01  &  5.51E-01  &  -0.1339 \\
                     &  (6.20E-02) & (1.13E-02) & (6.00E-03)  & (1.65E-02) & (8.89E-03) & \\
         ESO 12-G21  &  6.02E-02   &  1.14E-01  &  1.40E-01   &  1.64E-01  &  2.55E-01  &  -0.1241 \\
                     & (1.45E-02)  & (7.53E-03) & (5.46E-03)  & (6.74E-03) & (3.20E-03) & \\
          ESO 33-G2  &  8.98E-02   &  1.89E-01  &  3.11E-01   &  3.56E-01  &  3.16E-01  &  -0.0620 \\
                     & (1.84E-02)  & (1.69E-02) & (7.48E-03)  & (9.98E-03) & (4.86E-03) & \\
          FSC 09104  &  5.98E-02   &  1.35E-01  &  2.75E-01   &  3.70E-01  &  $\cdots$  &   0.2929 \\
                     & (6.92E-03)  & (1.40E-02) & (4.12E-03)  & (4.78E-03) &            & \\
           IC 4329A  &  5.53E-01   &  9.81E-01  &  1.60E+00   &  1.91E+00  &  1.52E+00  &  -0.0520 \\
                     & (2.30E-02)  & (6.17E-02) & (3.99E-02)  & (4.48E-02) & (1.36E-02) & \\
            IC 5063  &  $\cdots$   &  $\cdots$  &  2.18E+00   &  3.02E+00  &  3.89E+00  & $\cdots$  \\
                     &             &            &  (3.19E-02) & (3.37E-02) & (6.16E-02) & \\
    IRAS 05189-2524  &  2.04E-01   &  3.95E-01  &  1.16E+00   &  2.01E+00  &  5.91E+00  &   0.3367 \\
                     & (1.35E-02)  & (3.12E-02) & (1.91E-02)  & (1.29E-01) & (1.32E-01) & \\
   IRASF 01475-0740  &  4.46E-02   &  1.61E-01  &  2.61E-01   &  4.37E-01  &  4.81E-01  &  -0.1827 \\
                     & (1.10E-02)  & (1.57E-02) & (1.03E-02)  & (2.00E-02) & (7.02E-03) & \\
   IRASF 03450+0055  &  1.39E-01   &  2.65E-01  &  3.47E-01   &  4.57E-01  &  3.87E-01  &  -0.1850 \\
                     & (8.77E-03)  & (1.48E-02) & (5.59E-03)  & (1.54E-02) & (1.22E-02) & \\
   IRASF 04385-0828  & 2.27E-01    &  2.38E-01  &  8.02E-01   &  1.07E+00  &  1.46E+00  &   0.7856 \\
                     & (1.45E-02)  & (1.85E-02) & (2.88E-02)  & (2.24E-02) & (2.50E-02) & \\
   IRASF 15480-0344  &  4.29E-02   &  1.53E-01  &  2.99E-01   &  4.22E-01  &  4.92E-01  &  -0.0815 \\
                     & (9.36E-03)  & (2.09E-02) & (1.86E-02)  & (5.74E-03) & (8.65E-03) & \\
       MCG+0-29-23   &  6.69E-02   & 1.01E-01   &  2.48E-01   &  3.83E-01  &  8.54E-01  &   0.0629 \\
                     & (2.04E-02)  & (6.78E-03) & (6.52E-03)  & (8.24E-03) & (1.75E-02) & \\
     MCG-03-34-064   &  1.75E-01   &  5.04E-01  &  1.39E+00   &  2.05E+00  &  2.95E+00  &   0.2911 \\
                     & (2.53E-02)  & (6.26E-02) & (5.68E-02)  & (1.39E-02) & (5.14E-02) & \\
       MCG-2-33-34   &  2.74E-02   &  6.44E-02  &  1.12E-01   &  1.59E-01  &  2.31E-01  &  -0.0590 \\
                     & (6.77E-03)  & (1.87E-02) & (1.90E-02)  & (3.64E-03) & (6.59E-03) & \\
        MCG-2-40-4   &  2.02E-01   &  3.27E-01  &  5.45E-01   &  6.78E-01  &  9.53E-01  &   0.1155 \\
                     & (1.89E-02)  & (1.62E-02) & (1.93E-02)  & (1.34E-02) & (1.19E-02) & \\
        MCG-2-8-39   &  3.12E-02   &  1.21E-01  &  2.53E-01   &  2.92E-01  &  2.22E-01  &  -0.0427 \\
                     & (8.95E-03)  & (1.09E-02) & (7.18E-03)  & (1.61E-02) & (3.33E-03) & \\
       MCG-3-34-63   &  1.13E-02   &  1.61E-02  &  2.31E-02   &  4.08E-02  &  1.01E-01  &  -0.0420 \\
                     & (8.81E-03)  & (5.04E-03) & (3.59E-03)  & (6.41E-03) & (3.51E-03) & \\
        MCG-3-58-7   &  1.47E-01   &  2.78E-01  &  4.76E-01   &  7.01E-01  &  9.38E-01  &  -0.0072 \\
                     & (8.75E-03)  & (1.13E-02) & (1.21E-02)  & (2.02E-02) & (1.45E-02) & \\
       MCG-5-13-17   &  5.27E-02   &  1.11E-01  &  2.10E-01   &  3.19E-01  &  4.19E-01  &   0.1221 \\
                     & (1.13E-02)  & (9.16E-03) & (4.28E-03)  & (4.38E-03) & (5.80E-03) & \\
       MCG-6-30-15   &  1.80E-01   &  3.34E-01  &  4.80E-01   &  6.45E-01  &  6.10E-01  &  -0.0771 \\
                     & (1.34E-02)  & (2.20E-02) & (8.65E-03)  & (1.05E-02) & (6.27E-03) & \\
           Mrk 1239  &  4.57E-01   &  7.29E-01  &  8.85E-01   &  1.03E+00  &  9.14E-01  &  -0.0910 \\
                     & (1.60E-02)  & (2.24E-02) & (2.70E-02)  & (3.31E-02) & (1.42E-02) & \\
              Mrk 3  &  1.02E-01   &  2.94E-01  &  1.25E+00   &  2.07E+00  &  2.36E+00  &   0.3461 \\
                     & (1.12E-02)  & (4.24E-02) & (7.25E-02)  & (2.26E-02) & (2.27E-02) & \\
            Mrk 334  &  4.81E-02   &  1.13E-01  &  2.52E-01   &  5.71E-01  &  1.36E+00  &   0.2496 \\
                     & (7.58E-03)  & (7.10E-03) & (9.27E-03)  & (2.39E-02) & (4.01E-02) & \\
            Mrk 335  &  1.44E-01   &  2.03E-01  &  2.53E-01   &  2.80E-01  &  2.62E-01  &  -0.1258 \\
                     & (1.28E-02)  & (1.28E-02) & (9.50E-03)  & (9.27E-03) & (2.92E-03) & \\
            Mrk 348  &  1.25E-01   &  1.88E-01  &  4.30E-01   &  5.67E-01  &  4.95E-01  &   0.3455 \\
                     & (1.41E-02)  & (1.69E-02) & (1.79E-02)  & (1.51E-02) & (5.85E-03) & \\
           Mrk 463E  &  2.47E-01   &  3.30E-01  &  8.39E-01   &  1.28E+00  &  1.54E+00  &   0.3582 \\
                     & (1.42E-02)  & (3.13E-02) & (2.08E-02)  & (1.53E-02) & (1.44E-02) & \\
            Mrk 471  &  7.44E-03   &  1.58E-02  &  2.69E-02   &  4.31E-02  &  9.11E-02  &   0.1419 \\
                     & (1.59E-03)  & (1.15E-03) & (2.13E-03)  & (3.01E-03) & (3.88E-03) & \\
            Mrk 477  &  2.44E-02   &  7.46E-02  &  2.31E-01   &  4.21E-01  &  6.18E-01  &   0.2195 \\
                     &  (3.23E-03) & (9.39E-03) & (1.34E-02)  & (5.48E-03) & (6.51E-03) &    \\
            Mrk 609  &  1.76E-02   &  4.52E-02  &  9.80E-02   &  1.80E-01  &  4.35E-01  &   0.3291 \\
                     &  (5.24E-03) & (2.70E-03) & (6.13E-03)  & (7.76E-03) & (9.05E-03) &     \\
              Mrk 6  &  1.45E-01   &  2.08E-01  &  3.72E-01   &  5.37E-01  &  4.98E-01  &  -0.0280 \\
                     &  (1.02E-02) & (1.73E-02) & (1.72E-02)  & (1.80E-02) & (8.54E-03) &  \\
            Mrk 622  &  7.80E-03   &  3.78E-02  &  1.06E-01   &  2.55E-01  &  5.51E-01  &   0.1071 \\
                     &  (1.54E-03) & (3.40E-03) & (6.61E-03)  & (8.40E-03) & (5.31E-03) &          \\
            Mrk 704  &  1.91E-01   &  3.54E-01  &  5.05E-01   &  5.22E-01  &  4.02E-01  &  -0.0163 \\
                     &  (2.05E-02) & (1.90E-02) & (9.12E-03)  & (1.85E-02) & (6.29E-03) &          \\
            Mrk 766  &  1.49E-01   &  3.02E-01  &  6.95E-01   &  1.02E+00  &  1.38E+00  &   0.1533 \\
                     &  (1.62E-02) & (2.41E-02) & (2.73E-02)  & (8.78E-03) & (2.56E-02) &          \\
             Mrk 79  &  1.77E-01   &  2.98E-01  &  4.66E-01   &  5.98E-01  &  6.91E-01  &  -0.0141 \\
                     &  (1.58E-02) & (1.47E-02) & (1.78E-02)  & (1.04E-02) & (9.23E-03) &          \\
            Mrk 817  &  1.23E-01   &  2.91E-01  &  4.79E-01   &  8.08E-01  &  1.13E+00  &  -0.0577 \\
                     &  (2.39E-02) & (2.17E-02) & (2.37E-02)  & (3.47E-02) & (1.31E-02) &          \\
            Mrk 883  &  5.80E-03   &  1.67E-02  &  4.49E-02   &  1.18E-01  &  3.03E-01  &   0.1741 \\
                     &  (1.26E-03) & (2.41E-03) & (1.62E-03)  & (7.17E-03) & (7.93E-03) &          \\
              Mrk 9  &  1.04E-01   &  1.92E-01  &  2.62E-01   &  3.42E-01  &  3.91E-01  &   0.0281 \\
                     &  (1.92E-02) & (1.21E-02) & (6.06E-03)  & (7.71E-03) & (7.69E-03) &          \\
            Mrk 938  &  1.15E-01   &  9.93E-02  &  4.99E-01   &  1.08E+00  &  4.02E+00  &   1.0488 \\
                     &  (3.16E-02) & (1.64E-02) & (1.48E-02)  & (6.44E-02) & (1.36E-01) &          \\
           NGC 1056  &  6.46E-02   &  1.11E-01  &  1.54E-01   &  2.33E-01  &  5.45E-01  &  -0.0326 \\
                     &  (2.09E-02) & (1.07E-02) & (1.90E-02)  & (6.18E-03) & (1.29E-02) &          \\
           NGC 1125  &  4.71E-02   &  5.65E-02  &  2.72E-01   &  4.95E-01  &  1.08E+00  &   0.9306 \\
                     &  (1.18E-02) & (1.26E-02) & (1.79E-02)  & (1.23E-02) & (2.18E-02) &          \\
         NGC 1143/4  &  7.79E-02   &  8.75E-02  &  1.85E-01   &  $\cdots$  &  $\cdots$  &   0.4689 \\
                     &  (2.40E-02) & (1.34E-02) & (1.20E-02)  &            &            &          \\
           NGC 1194  &  1.52E-01   &  1.18E-01  &  4.03E-01   &  4.10E-01  &  4.72E-01  &   0.9308 \\
                     &  (1.47E-02) & (1.45E-02) & (8.76E-03)  & (7.86E-03) & (5.43E-03) &          \\
           NGC 1241  &  3.15E-02   &  4.78E-02  &  1.05E-01   &  1.58E-01  &  2.98E-01  &   0.3753 \\
                     &  (1.64E-02) & (1.21E-02) & (5.17E-03)  & (5.66E-03) & (6.42E-03) &          \\
           NGC 1320  &  1.44E-01   &  3.11E-01  &  5.61E-01   &  7.84E-01  &  9.40E-01  &   0.0757 \\
                     &  (1.65E-02) & (2.21E-02) & (1.53E-02)  & (1.04E-02) & (1.24E-02) &          \\
           NGC 1365  &  7.29E-01   &  1.23E+00  &  2.69E+00   &  4.89E+00  &  1.23E+01  &   0.2823 \\
                     &  (1.26E-01) & (7.41E-02) & (1.28E-01)  & (1.18E-01) & (3.43E-01) &          \\
           NGC 1386  &  2.03E-01   &  2.89E-01  &  8.44E-01   &  1.03E+00  &  1.58E+00  &   0.5700 \\
                     &  (1.50E-02) & (3.41E-02) & (3.92E-02)  & (1.86E-02) & (2.96E-02) &          \\
           NGC 1667  &  7.56E-02   &  1.36E-01  &  2.24E-01   &  3.04E-01  &  6.67E-01  &   0.0048 \\
                     &  (3.61E-02) & (1.07E-02) & (1.15E-02)  & (1.66E-02) & (1.94E-02) &          \\
           NGC 2273  &  1.19E-01   &  2.10E-01  &  5.54E-01   &  8.58E-01  &  1.69E+00  &   0.3614 \\
                     &  (1.42E-02) & (1.60E-02) & (1.16E-02)  & (1.69E-02) & (3.03E-02) &          \\
           NGC 2622  &  8.54E-03   &  2.19E-02  &  5.41E-02   &  8.32E-02  &  1.02E-01  &   0.0899 \\
                     &  (8.38E-04) & (2.41E-03) & (3.07E-03)  & (1.41E-03) & (1.18E-03) &          \\
           NGC 2639  &  5.99E-02   &  5.75E-02  &  8.29E-02   &  1.20E-01  &  1.99E-01  &   0.2610 \\
                     &  (2.11E-02) & (1.51E-02) & (8.20E-03)  & (1.87E-02) & (1.12E-02) &          \\
           NGC 2992  &  1.89E-01   &  3.87E-01  &  8.14E-01   &  1.03E+00  &  1.50E+00  &   0.2296 \\
                     &  (2.67E-02) & (4.50E-02) & (2.86E-02)  & (1.21E-02) & (3.51E-02) &          \\
           NGC 3079  &  3.33E-01   &  2.38E-01  &  7.11E-01   &  7.11E-01  &  3.36E+00  &   1.0463 \\
                     &  (7.64E-02) & (3.93E-02) & (7.85E-02)  & (3.76E-02) & (1.45E-01) &          \\
           NGC 3081  &  5.28E-02   &  1.74E-01  &  5.11E-01   &  7.95E-01  &  1.09E+00  &   0.0953 \\
                     &  (6.55E-03) & (2.56E-02) & (3.69E-02)  & (5.34E-03) & (8.37E-03) &          \\
           NGC 3227  &  $\cdots$   &  $\cdots$  &  7.22E-01   &  1.22E+00  &  1.76E+00  & $\cdots$ \\
                     &             &            &  (2.88E-02) & (3.15E-02) & (3.96E-02) &          \\
           NGC 3511  &  6.37E-02   &  8.39E-02  &  1.37E-01   &  1.72E-01  &  3.46E-01  &   0.1253 \\
                     &  (2.21E-02) & (1.46E-02) & (1.44E-02)  & (8.91E-03) & (1.43E-02) &          \\
           NGC 3516  &  1.97E-01   &  3.11E-01  &  4.80E-01   &  6.97E-01  &  7.77E-01  &   0.0365 \\
                     &  (1.27E-02) & (2.37E-02) & (1.45E-02)  & (7.83E-03) & (1.28E-02) &          \\
           NGC 3660  &  $\cdots$   &  3.14E-02  &  4.48E-02   &  9.01E-02  &  1.70E-01  &   0.6161 \\
                     &             &  (7.34E-03)& (3.92E-03)  & (8.20E-03) & (5.21E-03) &          \\
           NGC 3786  &  2.69E-02   &  4.85E-02  &  8.95E-02   &  1.43E-01  &  3.00E-01  &   0.1110 \\
                     &  (3.23E-03) & (4.80E-03) & (4.20E-03)  & (4.45E-03) & (6.09E-03) &          \\
           NGC 3982  &  6.43E-02   &  1.28E-01  &  2.03E-01   &  3.37E-01  &  6.56E-01  &   0.0014 \\
                     &  (2.51E-02) & (1.65E-02) & (5.24E-03)  & (1.23E-02) & (1.70E-02) &          \\
           NGC 4051  &  2.11E-01   &  4.61E-01  &  7.72E-01   &  1.10E+00  &  1.24E+00  &   0.0769 \\
                     &  (1.60E-02) & (2.23E-02) & (1.59E-02)  & (1.85E-02) & (1.24E-02) &          \\
           NGC 4151  &  6.80E-01   &  1.56E+00  &  3.16E+00   &  4.55E+00  &  3.64E+00  &  -0.0464 \\
                     &  (2.83E-02) & (1.22E-01) & (7.85E-02)  & (9.77E-02) & (3.89E-02) &          \\
            NGC 424  &  5.28E-01   &  9.06E-01  &  1.38E+00   &  1.63E+00  &  1.28E+00  &  -0.0307 \\
                     &  (2.50E-02) & (4.59E-02) & (4.90E-02)  & (3.80E-02) & (1.19E-02) &          \\
           NGC 4388  &  2.29E-01   &  2.90E-01  &  1.06E+00   &  1.80E+00  &  2.80E+00  &   0.7792 \\
                     &  (2.46E-02) & (6.73E-02) & (6.28E-02)  & (4.66E-02) & (4.83E-02) &          \\
           NGC 4501  &  1.15E-01   &  9.98E-02  &  1.16E-01   &  1.16E-01  &  1.38E-01  &   0.4831 \\
                     &  (1.56E-02) & (1.88E-02) & (5.89E-03)  & (1.82E-02) & (5.66E-03) &          \\
           NGC 4507  &  2.11E-01   &  4.71E-01  &  8.63E-01   &  1.28E+00  &  1.71E+00  &   0.0848 \\
                     &  (1.47E-02) & (4.04E-02) & (2.41E-02)  & (5.48E-03) & (3.23E-02) &          \\
           NGC 4579  &  1.31E-01   &  1.62E-01  &  1.24E-01   &  $\cdots$  &  $\cdots$  &  -0.3803 \\
                     &  (1.46E-02) & (4.21E-02) & (1.79E-02)  &            &            &          \\
           NGC 4593  &  2.22E-01   &  4.25E-01  &  5.43E-01   &  6.64E-01  &  8.13E-01  &  -0.0721 \\
                     &  (1.42E-02) & (1.50E-02) & (1.30E-02)  & (1.47E-02) & (1.18E-02) &          \\
           NGC 4594  &  2.43E-01   &  1.36E-01  &  8.21E-02   &  $\cdots$  &  $\cdots$  &  -0.3486 \\
                     &  (1.74E-02) & (7.20E-03) & (8.00E-03)  &            &            &          \\
           NGC 4602  &  4.78E-02   &  8.32E-02  &  1.20E-01   &  1.75E-01  &  3.65E-01  &  -0.0254 \\
                     &  (1.57E-02) & (1.19E-02) & (1.05E-02)  & (1.05E-02) & (1.30E-02) &          \\
      NGC 4922 NED01 &  1.61E-01   &  3.40E-01  &  8.56E-01   &  $\cdots$  &  $\cdots$  &   0.5302 \\
                     &   (3.98E-02)& (5.47E-02) & (6.60E-02)  &            &            &          \\
           NGC 4941  &  $\cdots$   &  $\cdots$  &  1.62E-01   &  2.99E-01  &  4.50E-01  & $\cdots$ \\
                     &             &            &  (1.13E-02) & (2.11E-02) & (1.30E-02) &          \\
           NGC 4968  &  1.03E-01   &  2.49E-01  &  6.21E-01   &  9.32E-01  &  1.02E+00  &   0.3205 \\
                     &  (2.13E-02) & (3.09E-02) & (2.95E-02)  & (2.00E-02) & (1.77E-02) &          \\
           NGC 5005  &  1.81E-01   &  1.59E-01  &  2.25E-01   &  2.85E-01  &  9.08E-01  &   0.6118 \\
                     &  (1.65E-02) & (3.35E-02) & (1.23E-02)  & (1.42E-02) & (3.66E-02) &          \\
            NGC 513  &  4.97E-02   &  9.63E-02  &  1.36E-01   &  1.98E-01  &  3.30E-01  &   0.1337 \\
                     &  (1.68E-02) & (1.06E-02) & (7.21E-03)  & (1.00E-02) & (1.17E-02) &          \\
           NGC 5135  &  1.51E-01   &  2.62E-01  &  6.70E-01   &  1.31E+00  &  3.03E+00  &   0.3971 \\
                     &  (3.08E-02) & (3.16E-02) & (3.61E-02)  & (3.52E-02) & (8.45E-02) &          \\
      NGC 5256 NED01 &  5.85E-02   &  7.62E-02  &  2.56E-01   &  4.95E-01  &  1.37E+00  &   0.5591 \\
                     &   (1.47E-02)& (1.69E-02) & (2.22E-02)  & (2.58E-02) & (2.67E-02) &          \\
           NGC 526A  &  $\cdots$   &  $\cdots$  &  3.08E-01   &  3.87E-01  &  2.55E-01  & $\cdots$ \\
                     &             &            &  (1.35E-02) & (1.23E-02) & (1.23E-02) &          \\
           NGC 5347  &  6.19E-02   &  2.10E-01  &  5.22E-01   &  7.86E-01  &  7.91E-01  &   0.0025 \\
                     &  (8.71E-03) & (1.92E-02) & (1.22E-02)  & (1.78E-02) & (1.03E-02) &          \\
           NGC 5506  &  9.33E-01   &  6.88E-01  &  2.18E+00   &  2.86E+00  &  4.05E+00  &   0.7640 \\
                     &  (2.17E-02) & (1.06E-01) & (6.61E-02)  & (5.59E-02) & (6.38E-02) &          \\
           NGC 5929  &  1.35E-02   &  2.08E-02  &  2.89E-02   &  5.84E-02  &  1.13E-01  &   0.0243 \\
                     &  (7.60E-03) & (7.60E-03) & (2.66E-03)  & (3.46E-03) & (3.62E-03) &          \\
           NGC 5953  &  1.20E-01   &  1.87E-01  &  3.16E-01   &  5.14E-01  &  1.27E+00  &   0.0941 \\
                     &  (3.57E-02) & (1.81E-02) & (3.25E-02)  & (1.47E-02) & (3.37E-02) &          \\
           NGC 6810  &  2.02E-01   &  4.13E-01  &  9.59E-01   &  2.13E+00  &  3.33E+00  &   0.2068 \\
                     &  (4.37E-02) & (2.65E-02) & (2.85E-02)  & (2.86E-02) & (7.18E-02) &          \\
           NGC 6860  &  1.64E-01   &  2.48E-01  &  3.25E-01   &  3.87E-01  &  3.36E-01  &  -0.0012 \\
                     &  (1.28E-02) & (5.12E-03) & (1.26E-02)  & (1.16E-02) & (4.73E-03) &          \\
           NGC 6890  &  8.02E-02   &  1.49E-01  &  2.62E-01   &  3.98E-01  &  5.79E-01  &  -0.0208 \\
                     &  (1.55E-02) & (1.10E-02) & (8.16E-03)  & (1.08E-02) & (9.52E-03) &          \\
           NGC 7130  &  9.29E-02   &  2.05E-01  &  5.38E-01   &  1.08E+00  &  2.47E+00  &   0.2710 \\
                     &  (1.67E-02) & (1.62E-02) & (2.21E-02)  & (3.30E-02) & (6.89E-02) &          \\
           NGC 7172  &  $\cdots$   & $\cdots$   &  4.20E-01   &  3.89E-01  &  9.81E-01  & $\cdots$ \\
                     &             &            &  (3.05E-02) & (2.41E-02) & (2.94E-02) &          \\
           NGC 7213  &  1.79E-01   &  4.14E-01  &  4.87E-01   &  7.33E-01  &  5.77E-01  &  -0.2470 \\
                     &  (1.62E-02) & (2.80E-02) & (1.59E-02)  & (2.00E-02) & (1.68E-02) &          \\
           NGC 7314  &  $\cdots$   & $\cdots$   &  2.23E-01   &  2.71E-01  &  4.36E-01  & $\cdots$ \\
                     &             &            &  (2.10E-02) & (1.46E-02) & (7.01E-03) &          \\
           NGC 7469  &  3.31E-01   &  7.67E-01  &  1.65E+00   &  3.28E+00  &  6.11E+00  &   0.1045 \\
                     &  (4.74E-02) & (4.89E-02) & (4.61E-02)  & (1.04E-01) & (1.67E-01) &          \\
           NGC 7496  &  4.71E-02   &  1.37E-01  &  3.77E-01   &  8.96E-01  &  2.14E+00  &   0.1744 \\
                     &  (1.41E-02) & (1.53E-02) & (1.85E-02)  & (2.54E-02) & (5.02E-02) &          \\
           NGC 7582  &  6.34E-01   &  5.79E-01  &  1.97E+00   &  3.52E+00  &  9.33E+00  &   0.8284 \\
                     &  (5.33E-02) & (7.27E-02) & (8.92E-02)  & (1.22E-01) & (2.74E-01) &          \\
           NGC 7590  &  4.44E-02   &  6.55E-02  &  8.54E-02   &  1.20E-01  &  2.46E-01  &  -0.0658 \\
                     &  (1.32E-02) & (8.10E-03) & (6.82E-03)  & (5.80E-03) & (1.17E-02) &          \\
           NGC 7603  &  2.39E-01   &  3.31E-01  &  2.98E-01   &  3.06E-01  &  3.12E-01  &  -0.0630 \\
                     &  (1.25E-02) & (6.28E-03) & (1.41E-02)  & (1.11E-02) & (6.83E-03) &          \\
           NGC 7674  &  1.89E-01   &  3.89E-01  &  9.24E-01   &  1.37E+00  &  1.83E+00  &   0.2811 \\
                     &  (2.60E-02) & (3.44E-02) & (3.37E-02)  & (2.31E-02) & (1.57E-02) &          \\
            NGC 788  &  3.01E-02   &  8.23E-02  &  2.16E-01   &  3.01E-01  &  3.47E-01  &   0.1234 \\
                     &  (3.30E-03) & (1.05E-02) & (7.87E-03)  & (1.34E-03) & (3.59E-03) &          \\
            NGC 931  &  2.22E-01   &  3.99E-01  &  6.56E-01   &  8.79E-01  &  9.29E-01  &   0.0166 \\
                     &  (1.81E-02) & (2.04E-02) & (2.67E-02)  & (1.40E-02) & (9.02E-03) &          \\
    SDSS J1039+6430  &  6.74E-03   &  1.33E-02  &  2.72E-02   &  $\cdots$  & $\cdots$   &  -0.0103 \\
                     &  (2.00E-04) & (1.17E-03) & (7.75E-04)  &            &            &          \\
    SDSS J1641+3858  &  3.75E-03   &  2.13E-02  &  $\cdots$   & $\cdots$   & $\cdots$   &  -0.0473 \\
                     &  (1.02E-03) & (2.20E-03) &             &            &            &          \\
        TOL1238-364  &  1.18E-01   &  3.36E-01  &  9.20E-01   &  1.67E+00  &  2.16E+00  &   0.2025 \\
                     &  (1.90E-02) & (3.82E-02) & (3.69E-02)  & (1.58E-02) & (3.93E-02) &          \\
    UGC 11680 NED01  &  3.62E-02   &  8.87E-02  &  1.33E-01   &  1.80E-01  &  2.11E-01  &   0.0946 \\
                     &  (6.71E-03) & (1.54E-02) & (7.37E-03)  & (1.12E-02) & (6.44E-03) &          \\
          UGC 12138  &  2.43E-02   &  5.56E-02  &  1.11E-01   &  1.79E-01  &  3.03E-01  &   0.2383 \\
                     &  (2.07E-03) & (5.07E-03) & (5.79E-03)  & (3.68E-03) & (1.09E-02) &          \\
           UGC 7064  &  3.06E-02   &  8.51E-02  &  1.52E-01   &  2.31E-01  &  3.28E-01  &   0.1045 \\
                     &  (1.36E-02) & (6.15E-03) & (6.79E-03)  & (1.16E-02) & (5.85E-03) &          \\
             UM 146  &  6.76E-03   &  1.42E-02  &  3.22E-02   &  5.29E-02  &  9.72E-02  &   0.2348 \\
                     &  (8.04E-04) & (1.05E-03) & (1.62E-03)  & (3.55E-03) & (5.13E-04) &          \\
WIR-IRAS 23060+0505  &  1.01E-01   &  1.58E-01  &  3.31E-01   &  4.53E-01  &  7.51E-01  &   0.3170 \\
                     &  (6.50E-03) & (9.47E-03) & (4.01E-03)  & (1.27E-02) & (1.46E-02) &          \\
    \enddata
    \tablecomments{Col. (1): Galaxy name; Col. (2-6): Continuum flux density
      as measured on the spectrum at 5.5, 10., 14.7, 20, and 30$\mum$ in
      Jansky; Col. (7): Observed optical depth (apparent) at 9.7$\mum$;
      1~$\sigma$ errors are given in parenthesis.}
    \label{tab:contflux}
  \end{deluxetable}
  \clearpage

  \LongTables
  \begin{deluxetable}{lccccc}
    \tabletypesize{\tiny}
    \tablewidth{0pt}
    \tablecolumns{6}
    \tablecaption{Starburst-to-AGN Flux Density Ratios}
    \tablehead{\colhead{Galaxy Name} &
      \multicolumn{5}{c}{Starburst-to-AGN Flux Density Ratio} \\
      \colhead{} &
      \colhead{5.5$\mum$} &
      \colhead{10$\mum$} &
      \colhead{14.7$\mum$} &
      \colhead{20$\mum$} &
      \colhead{30$\mum$} \\
      \colhead{(1)} &
      \colhead{(2)} &
      \colhead{(3)} &
      \colhead{(4)} &
      \colhead{(5)} &
      \colhead{(6)} \\
    }
    \startdata
3C 321                   &  0.0263     &  0.0400      &  0.0207        &  0.0354       &  0.0640 \\
CGCG 381-051              &  $\cdots$   &  0.2681      &  0.5661        &  0.9087       &  $\cdots$ \\
ESO 12-G21               &  0.3020     &  0.7335      &  13.4681       &  $\cdots$     &  $\cdots$ \\
ESO 33-G2                &  0.0291     &  0.0359      &  0.0622        &  0.1534       &  0.9593 \\
IC 4329A                 &  0.0043     &  0.0082      &  0.0145        &  0.0317       &  0.1481 \\
IRAS 05189-2524          &  0.0355     &  0.0640      &  0.0610        &  0.0925       &  0.1097 \\
IRASF 01475-0740         &  0.0702     &  0.0835      &  0.1556        &  0.2870       &  1.7003 \\
IRASF 04385-0828         &  0.0253     &  0.0887      &  0.0715        &  0.1456       &  0.4536 \\
IRASF 15480-0344         &  0.0679     &  0.0696      &  0.1084        &  0.2052       &  0.9631 \\
MCG+0-29-23              &  1.4460     &  11.5036     &  $\cdots$      &  $\cdots$     &  $\cdots$ \\
MCG-03-34-064            &  0.0247     &  0.0309      &  0.0323        &  0.0571       &  0.1434 \\
MCG-2-33-34              &  0.1739     &  0.3531      &  0.6773        &  2.3617       &  $\cdots$ \\
MCG-2-40-4               &  0.0613     &  0.1260      &  0.2427        &  0.6416       &  17.9313 \\
MCG-3-34-63              &  0.3026     &  38.4053     &  $\cdots$      &  $\cdots$     &  $\cdots$ \\
MCG-3-58-7               &  0.0501     &  0.0891      &  0.1635        &  0.3254       &  1.5569 \\
MCG-5-13-17              &  0.1574     &  0.3035      &  0.5114        &  1.3713       &  $\cdots$ \\
MCG-6-30-15              &  0.0171     &  0.0288      &  0.0588        &  0.1249       &  0.6378 \\
Mrk 1239                 &  0.0138     &  0.0281      &  0.0706        &  0.1727       &  1.1758 \\
Mrk 3                    &  0.0060     &  0.0077      &  0.0051        &  0.0077       &  0.0233 \\
Mrk 334                  &  0.4212     &  0.6677      &  1.0960        &  1.4361       &  4.5979 \\
Mrk 348                  &  0.0395     &  0.0754      &  0.0971        &  0.2081       &  1.8993 \\
Mrk 463E                 &  0.0124     &  0.0329      &  0.0364        &  0.0627       &  0.1964 \\
Mrk 471                  &  0.6422     &  1.7036      &  $\cdots$      &  $\cdots$     &  $\cdots$ \\
Mrk 477                  &  0.0695     &  0.0817      &  0.0705        &  0.1061       &  0.2781 \\
Mrk 6                    &  0.0170     &  0.0411      &  0.0668        &  0.1237       &  0.6543 \\
Mrk 609                  &  1.3144     &  2.3839      &  25.1028       &  $\cdots$     &  $\cdots$ \\
Mrk 622                  &  0.5186     &  0.3268      &  0.3328        &  0.3661       &  0.7087 \\
Mrk 766                  &  0.0574     &  0.0914      &  0.1203        &  0.2250       &  0.7952 \\
Mrk 79                   &  0.0211     &  0.0429      &  0.0809        &  0.1754       &  0.7295 \\
Mrk 817                  &  0.0417     &  0.0625      &  0.1186        &  0.1911       &  0.6321 \\
Mrk 883                  &  0.5226     &  0.8595      &  0.7567        &  0.6741       &  1.1640 \\
Mrk 9                    &  0.0627     &  0.0951      &  0.2400        &  0.6527       &  $\cdots$ \\
Mrk 938                  &  0.7001     &  $\cdots$    &  6.2222        &  $\cdots$     &  11.4720 \\
NGC 1125                 &  0.3060     &  1.8193      &  0.6179        &  1.1437       &  4.2422 \\
NGC 1143/4               &  0.6227     &  $\cdots$    &  $\cdots$      &  $\cdots$     &  $\cdots$ \\
NGC 1241                 &  0.2527     &  1.2900      &  3.9115        &  $\cdots$     &  $\cdots$ \\
NGC 1320                 &  0.0667     &  0.0964      &  0.1615        &  0.3454       &  2.5940 \\
NGC 1365                 &  0.8418     &  3.5882      &  2280.9476     &  $\cdots$     &  $\cdots$ \\
NGC 1386                 &  0.0528     &  0.1621      &  0.1460        &  0.3804       &  1.4503 \\
NGC 2273                 &  0.2300     &  0.4914      &  0.5444        &  1.4474       &  88.6744 \\
NGC 2622                 &  0.1445     &  0.2242      &  0.2628        &  0.5276       &  13.8732 \\
NGC 2639                 &  0.3437     &  31.3315     &  $\cdots$      &  $\cdots$     &  $\cdots$ \\
NGC 2992                 &  0.2337     &  0.4174      &  0.6511        &  3.4637       &  $\cdots$ \\
NGC 3079                 &  4.6747     &  $\cdots$    &  $\cdots$      &  $\cdots$     &  $\cdots$ \\
NGC 3081                 &  0.0442     &  0.0485      &  0.0480        &  0.0785       &  0.2137 \\
NGC 3227                 &  $\cdots$   &  $\cdots$    &  0.4713        &  0.9641       &  $\cdots$ \\
NGC 3511                 &  2.6708     &  $\cdots$    &  $\cdots$      &  $\cdots$     &  $\cdots$ \\
NGC 3516                 &  0.0170     &  0.0389      &  0.0758        &  0.1423       &  0.5974 \\
NGC 3660                 &  $\cdots$   &  1.1048      &  $\cdots$      &  $\cdots$     &  $\cdots$ \\
NGC 3786                 &  0.2673     &  0.6172      &  1.4630        &  11.1716      &  $\cdots$ \\
NGC 3982                 &  2.6710     &  9.3812      &  $\cdots$      &  $\cdots$     &  $\cdots$ \\
NGC 4051                 &  0.0702     &  0.1056      &  0.1872        &  0.4395       &  7.1327 \\
NGC 4151                 &  0.0129     &  0.0186      &  0.0263        &  0.0497       &  0.2452 \\
NGC 424                  &  0.0102     &  0.0200      &  0.0390        &  0.0885       &  0.5310 \\
NGC 4388                 &  0.1237     &  0.5162      &  0.3060        &  0.5262       &  2.9754 \\
NGC 4501                 &  0.1361     &  1.2925      &  $\cdots$      &  $\cdots$     &  $\cdots$ \\
NGC 4507                 &  0.0242     &  0.0372      &  0.0588        &  0.1080       &  0.3233 \\
NGC 4579                 &  0.0539     &  0.1936      &  $\cdots$      &  $\cdots$     &  $\cdots$ \\
NGC 4593                 &  0.0346     &  0.0630      &  0.1552        &  0.3887       &  3.2523 \\
NGC 4594                 &  0.0094     &  0.0609      &  $\cdots$      &  $\cdots$     &  $\cdots$ \\
NGC 4602                 &  0.8372     &  2.8868      &  $\cdots$      &  $\cdots$     &  $\cdots$ \\
NGC 4922 NED01           &  0.1114     &  0.2474      &  $\cdots$      &  $\cdots$     &  $\cdots$ \\
NGC 4941                 &  $\cdots$   &  $\cdots$    &  0.8976        &  1.4715       &  $\cdots$ \\
NGC 4968                 &  0.1301     &  0.1651      &  0.1894        &  0.3638       &  4.7134 \\
NGC 5005                 &  0.1619     &  2.3674      &  $\cdots$      &  $\cdots$     &  $\cdots$ \\
NGC 513                  &  0.9011     &  1.1506      &  $\cdots$      &  $\cdots$     &  $\cdots$ \\
NGC 5135                 &  0.7242     &  3.4740      &  8.6441        &  $\cdots$     &  $\cdots$ \\
NGC 5256 NED01           &  1.1450     &  65.3742     &  8.9031        &  $\cdots$     &  $\cdots$ \\
NGC 5347                 &  0.0552     &  0.0598      &  0.0665        &  0.1229       &  0.5838 \\
NGC 5506                 &  0.0250     &  0.1339      &  0.1109        &  0.2495       &  0.9290 \\
NGC 5929                 &  0.5391     &  $\cdots$    &  $\cdots$      &  $\cdots$     &  $\cdots$ \\
NGC 5953                 &  4.1648     &  $\cdots$    &  $\cdots$      &  $\cdots$     &  $\cdots$ \\
NGC 6810                 &  1.0366     &  1.9786      &  4.2739        &  17.9650      &  $\cdots$ \\
NGC 6860                 &  0.0602     &  0.1206      &  0.3297        &  1.0168       &  $\cdots$ \\
NGC 6890                 &  0.2445     &  0.4524      &  0.9600        &  5.6312       &  $\cdots$ \\
NGC 7130                 &  1.0761     &  2.3340      &  2.4662        &  10.1020      &  $\cdots$ \\
NGC 7213                 &  0.0530     &  0.0914      &  0.2545        &  0.5246       &  $\cdots$ \\
NGC 7314                 &  $\cdots$   &  $\cdots$    &  0.4652        &  1.9107       &  $\cdots$ \\
NGC 7469                 &  0.3845     &  0.5837      &  0.9244        &  1.6895       &  $\cdots$ \\
NGC 7496                 &  1.3900     &  1.0345      &  0.9555        &  1.2744       &  3.7034 \\
NGC 7582                 &  0.3118     &  9.1000      &  2.2330        &  536.9538     &  $\cdots$ \\
NGC 7590                 &  0.6934     &  $\cdots$    &  $\cdots$      &  $\cdots$     &  $\cdots$ \\
NGC 7603                 &  0.0458     &  0.1133      &  0.4841        &  3.8650       &  $\cdots$ \\
NGC 7674                 &  0.1407     &  0.2002      &  0.2614        &  0.5388       &  7.2992 \\
NGC 788                  &  0.0274     &  0.0361      &  0.0380        &  0.0728       &  0.2437 \\
NGC 931                  &  0.0326     &  0.0600      &  0.1093        &  0.2347       &  1.4468 \\
SDSS J1039+6430          &  0.0114     &  0.0188      &  0.0268        &  $\cdots$     &  $\cdots$ \\
TOL 1238-364             &  0.3450     &  0.3095      &  0.3202        &  0.5216       &  8.6584 \\
UGC 11680 NED01          &  0.1504     &  0.1820      &  0.4536        &  1.4949       &  $\cdots$ \\
UGC 12138                &  0.1290     &  0.2047      &  0.3179        &  0.6320       &  3.1993 \\
UGC 7064                 &  0.6323     &  0.4803      &  1.2807        &  276.5345     &  $\cdots$ \\
UM 146                   &  0.0639     &  0.1100      &  0.1527        &  0.2621       &  0.5764 \\
WIR-IRAS 23060+0505      &  0.0113     &  0.0248      &  0.0338        &  0.0665       &  0.1457 \\
    \enddata
    \tablecomments{Col. (1): Galaxy name; Col. (2-6): Starburst-to-AGN flux
      density ratio measured at 5.5, 10, 14.7, 20, and 30$\mum$; a missing
      ratio implies either the spectrum was incomplete at that wavelength or
      the AGN contribution was negative after starburst subtraction. A very
      large ratio indicates very weak AGN contribution, these typically occur
      at or beyond 14.7$\mum$ in a few galaxies. See Figure~\ref{fig:lum1} for
      a visual representation of these ratios, and Table~\ref{tab:objdata} for
      aperture size in parsecs. Even in nearby AGN {\it Spitzer} spectra
      sample regions of size $\sim$ 1$\kpc$. See Section~5 for discussion of
      the starburst subtraction process and results.}
    \label{tab:sbratio}
  \end{deluxetable}
  \clearpage


  \clearpage

  \clearpage
  
  \clearpage

  \clearpage

  \clearpage

  \clearpage

  \clearpage

  \clearpage

  \clearpage

  \clearpage

  \clearpage


\begin{thebibliography}{45}
\expandafter\ifx\csname natexlab\endcsname\relax\def\natexlab#1{#1}\fi

\bibitem[{{Alonso-Herrero} {et~al.}(2001){Alonso-Herrero}, {Quillen},
  {Simpson}, {Efstathiou}, \& {Ward}}]{2001AJ....121.1369A}
{Alonso-Herrero}, A., {Quillen}, A.~C., {Simpson}, C., {Efstathiou}, A., \&
  {Ward}, M.~J. 2001, \aj, 121, 1369

\bibitem[{{Antonucci}(1993)}]{1993ARA&A..31..473A}
{Antonucci}, R. 1993, \araa, 31, 473

\bibitem[{{Antonucci} \& {Miller}(1985)}]{1985ApJ...297..621A}
{Antonucci}, R.~R.~J., \& {Miller}, J.~S. 1985, \apj, 297, 621

\bibitem[{{Baldwin} {et~al.}(1981){Baldwin}, {Phillips}, \&
  {Terlevich}}]{1981PASP...93....5B}
{Baldwin}, J.~A., {Phillips}, M.~M., \& {Terlevich}, R. 1981, \pasp, 93, 5

\bibitem[{{Brandl} {et~al.}(2006){Brandl}, {Bernard-Salas}, {Spoon}, {Devost},
  {Sloan}, {Guilles}, {Wu}, {Houck}, {Weedman}, {Armus}, {Appleton}, {Soifer},
  {Charmandaris}, {Hao}, {Higdon}, \& {Herter}}]{2006ApJ...653.1129B}
{Brandl}, B.~R., \etal 2006, \apj, 653, 1129

\bibitem[{{Buchanan} {et~al.}(2006){Buchanan}, {Gallimore}, {O'Dea}, {Baum},
  {Axon}, {Robinson}, {Elitzur}, \& {Elvis}}]{2006AJ....132..401B}
{Buchanan}, C.~L., {Gallimore}, J.~F., {O'Dea}, C.~P., {Baum}, S.~A., {Axon},
  D.~J., {Robinson}, A., {Elitzur}, M., \& {Elvis}, M. 2006, \aj, 132, 401

\bibitem[{{Clavel} {et~al.}(2000){Clavel}, {Schulz}, {Altieri}, {Barr},
  {Claes}, {Heras}, {Leech}, {Metcalfe}, \& {Salama}}]{2000A&A...357..839C}
{Clavel}, J., \etal 2000, \aap, 357, 839

\bibitem[{{Deo} {et~al.}(2007){Deo}, {Crenshaw}, {Kraemer}, {Dietrich},
  {Elitzur}, {Teplitz}, \& {Turner}}]{2007ApJ...671..124D}
{Deo}, R.~P., {Crenshaw}, D.~M., {Kraemer}, S.~B., {Dietrich}, M., {Elitzur},
  M., {Teplitz}, H., \& {Turner}, T.~J. 2007, \apj, 671, 124

\bibitem[{{Edelson} \& {Malkan}(1986)}]{1986ApJ...308...59E}
{Edelson}, R.~A., \& {Malkan}, M.~A. 1986, \apj, 308, 59

\bibitem[{{Gallagher} {et~al.}(2007){Gallagher}, {Richards}, {Lacy}, {Hines},
  {Elitzur}, \& {Storrie-Lombardi}}]{2007ApJ...661...30G}
{Gallagher}, S.~C., {Richards}, G.~T., {Lacy}, M., {Hines}, D.~C., {Elitzur},
  M., \& {Storrie-Lombardi}, L.~J. 2007, \apj, 661, 30

\bibitem[{{Genzel} {et~al.}(1998){Genzel}, {Lutz}, {Sturm}, {Egami}, {Kunze},
  {Moorwood}, {Rigopoulou}, {Spoon}, {Sternberg}, {Tacconi-Garman}, {Tacconi},
  \& {Thatte}}]{1998ApJ...498..579G}
{Genzel}, R., \etal 1998, \apj, 498, 579

\bibitem[{{Haas} {et~al.}(2007){Haas}, {Siebenmorgen}, {Pantin}, {Horst},
  {Smette}, {K{\"a}ufl}, {Lagage}, \& {Chini}}]{2007A&A...473..369H}
{Haas}, M., {Siebenmorgen}, R., {Pantin}, E., {Horst}, H., {Smette}, A.,
  {K{\"a}ufl}, H.-U., {Lagage}, P.-O., \& {Chini}, R. 2007, \aap, 473, 369

\bibitem[{{Haas} {et~al.}(2003){Haas}, {Klaas}, {M{\"u}ller}, {Bertoldi},
  {Camenzind}, {Chini}, {Krause}, {Lemke}, {Meisenheimer}, {Richards}, \&
  {Wilkes}}]{2003A&A...402...87H}
{Haas}, M., \etal 2003, \aap, 402, 87

\bibitem[{{Hao} {et~al.}(2007){Hao}, {Weedman}, {Spoon}, {Marshall},
  {Levenson}, {Elitzur}, \& {Houck}}]{2007ApJ...655L..77H}
{Hao}, L., {Weedman}, D.~W., {Spoon}, H.~W.~W., {Marshall}, J.~A., {Levenson},
  N.~A., {Elitzur}, M., \& {Houck}, J.~R. 2007, \apjl, 655, L77

\bibitem[{{Higdon} {et~al.}(2004){Higdon}, {Devost}, {Higdon}, {Brandl},
  {Houck}, {Hall}, {Barry}, {Charmandaris}, {Smith}, {Sloan}, \&
  {Green}}]{2004PASP..116..975H}
{Higdon}, S.~J.~U., \etal 2004, \pasp, 116, 975

\bibitem[{{Houck} {et~al.}(2004){Houck}, {Roellig}, {van Cleve}, {Forrest},
  {Herter}, {Lawrence}, {Matthews}, {Reitsema}, {Soifer}, {Watson}, {Weedman},
  {Huisjen}, {Troeltzsch}, {Barry}, {Bernard-Salas}, {Blacken}, {Brandl},
  {Charmandaris}, {Devost}, {Gull}, {Hall}, {Henderson}, {Higdon}, {Pirger},
  {Schoenwald}, {Sloan}, {Uchida}, {Appleton}, {Armus}, {Burgdorf},
  {Fajardo-Acosta}, {Grillmair}, {Ingalls}, {Morris}, \&
  {Teplitz}}]{2004ApJS..154...18H}
{Houck}, J.~R., \etal 2004, \apjs, 154, 18

\bibitem[{{Imanishi} \& {Alonso-Herrero}(2004)}]{2004ApJ...614..122I}
{Imanishi}, M., \& {Alonso-Herrero}, A. 2004, \apj, 614, 122

\bibitem[{{Keel}(1980)}]{1980AJ.....85..198K}
{Keel}, W.~C. 1980, \aj, 85, 198

\bibitem[{{Klaas} {et~al.}(2001){Klaas}, {Haas}, {M{\"u}ller}, {Chini},
  {Schulz}, {Coulson}, {Hippelein}, {Wilke}, {Albrecht}, \&
  {Lemke}}]{2001A&A...379..823K}
{Klaas}, U., \etal 2001, \aap, 379, 823

\bibitem[{{Laurent} {et~al.}(2000){Laurent}, {Mirabel}, {Charmandaris},
  {Gallais}, {Madden}, {Sauvage}, {Vigroux}, \&
  {Cesarsky}}]{2000A&A...359..887L}
{Laurent}, O., {Mirabel}, I.~F., {Charmandaris}, V., {Gallais}, P., {Madden},
  S.~C., {Sauvage}, M., {Vigroux}, L., \& {Cesarsky}, C. 2000, \aap, 359, 887

\bibitem[{{Lumsden} {et~al.}(2004){Lumsden}, {Alexander}, \&
  {Hough}}]{2004MNRAS.348.1451L}
{Lumsden}, S.~L., {Alexander}, D.~M., \& {Hough}, J.~H. 2004, \mnras, 348, 1451

\bibitem[{{Maiolino} \& {Rieke}(1995)}]{1995ApJ...454...95M}
{Maiolino}, R., \& {Rieke}, G.~H. 1995, \apj, 454, 95

\bibitem[{{Mel{\'e}ndez} {et~al.}(2008{\natexlab{a}}){Mel{\'e}ndez}, {Kraemer},
  {Armentrout}, {Deo}, {Crenshaw}, {Schmitt}, {Mushotzky}, {Tueller},
  {Markwardt}, \& {Winter}}]{2008ApJ...682...94M}
{Mel{\'e}ndez}, M., \etal 2008{\natexlab{a}}, \apj, 682, 94

\bibitem[{{Mel{\'e}ndez} {et~al.}(2008{\natexlab{b}}){Mel{\'e}ndez}, {Kraemer},
  {Schmitt}, {Crenshaw}, {Deo}, {Mushotzky}, \&
  {Bruhweiler}}]{2008ApJ...689...95M}
{Mel{\'e}ndez}, M., {Kraemer}, S.~B., {Schmitt}, H.~R., {Crenshaw}, D.~M.,
  {Deo}, R.~P., {Mushotzky}, R.~F., \& {Bruhweiler}, F.~C. 2008{\natexlab{b}},
  \apj, 689, 95

\bibitem[{{Mushotzky} {et~al.}(2008){Mushotzky}, {Winter}, {McIntosh}, \&
  {Tueller}}]{2008arXiv0807.4695M}
{Mushotzky}, R.~F., {Winter}, L.~M., {McIntosh}, D.~H., \& {Tueller}, J. 2008,
  ArXiv e-prints:0807.4695M

\bibitem[{{Nagao} {et~al.}(2004){Nagao}, {Kawabata}, {Murayama}, {Ohyama},
  {Taniguchi}, {Sumiya}, \& {Sasaki}}]{2004AJ....128..109N}
{Nagao}, T., {Kawabata}, K.~S., {Murayama}, T., {Ohyama}, Y., {Taniguchi}, Y.,
  {Sumiya}, R., \& {Sasaki}, S.~S. 2004, \aj, 128, 109

\bibitem[{{Nardini} {et~al.}(2008){Nardini}, {Risaliti}, {Salvati}, {Sani},
  {Imanishi}, {Marconi}, \& {Maiolino}}]{2008MNRAS.385L.130N}
{Nardini}, E., {Risaliti}, G., {Salvati}, M., {Sani}, E., {Imanishi}, M.,
  {Marconi}, A., \& {Maiolino}, R. 2008, \mnras, 385, L130

\bibitem[{{Peeters} {et~al.}(2004){Peeters}, {Spoon}, \&
  {Tielens}}]{2004ApJ...613..986P}
{Peeters}, E., {Spoon}, H.~W.~W., \& {Tielens}, A.~G.~G.~M. 2004, \apj, 613,
  986

\bibitem[{{Richards} {et~al.}(2006){Richards}, {Lacy}, {Storrie-Lombardi},
  {Hall}, {Gallagher}, {Hines}, {Fan}, {Papovich}, {Vanden Berk}, {Trammell},
  {Schneider}, {Vestergaard}, {York}, {Jester}, {Anderson}, {Budav{\'a}ri}, \&
  {Szalay}}]{2006ApJS..166..470R}
{Richards}, G.~T., \etal 2006, \apjs, 166, 470

\bibitem[{{Rodr{\'{\i}}guez-Ardila} \& {Mazzalay}(2006)}]{2006MNRAS.367L..57R}
{Rodr{\'{\i}}guez-Ardila}, A., \& {Mazzalay}, X. 2006, \mnras, 367, L57

\bibitem[{{Rodriguez Espinosa} {et~al.}(1996){Rodriguez Espinosa}, {Perez
  Garcia}, {Lemke}, \& {Meisenheimer}}]{1996A&A...315L.129R}
{Rodriguez Espinosa}, J.~M., {Perez Garcia}, A.~M., {Lemke}, D., \&
  {Meisenheimer}, K. 1996, \aap, 315, L129

\bibitem[{{Schweitzer} {et~al.}(2006){Schweitzer}, {Lutz}, {Sturm}, {Contursi},
  {Tacconi}, {Lehnert}, {Dasyra}, {Genzel}, {Veilleux}, {Rupke}, {Kim},
  {Baker}, {Netzer}, {Sternberg}, {Mazzarella}, \&
  {Lord}}]{2006ApJ...649...79S}
{Schweitzer}, M., \etal 2006, \apj, 649, 79

\bibitem[{{Seyfert}(1943)}]{1943ApJ....97...28S}
{Seyfert}, C.~K. 1943, \apj, 97, 28

\bibitem[{{Smith} {et~al.}(2007{\natexlab{a}}){Smith}, {Armus}, {Dale},
  {Roussel}, {Sheth}, {Buckalew}, {Jarrett}, {Helou}, \&
  {Kennicutt}}]{2007PASP..119.1133S}
{Smith}, J.~D.~T., \etal 2007{\natexlab{a}}, \pasp, 119, 1133

\bibitem[{{Smith} {et~al.}(2007{\natexlab{b}}){Smith}, {Draine}, {Dale},
  {Moustakas}, {Kennicutt}, {Helou}, {Armus}, {Roussel}, {Sheth}, {Bendo},
  {Buckalew}, {Calzetti}, {Engelbracht}, {Gordon}, {Hollenbach}, {Li},
  {Malhotra}, {Murphy}, \& {Walter}}]{2007ApJ...656..770S}
{Smith}, J.~D.~T., \etal 2007{\natexlab{b}}, \apj, 656, 770

\bibitem[{{Spoon} {et~al.}(2007){Spoon}, {Marshall}, {Houck}, {Elitzur}, {Hao},
  {Armus}, {Brandl}, \& {Charmandaris}}]{2007ApJ...654L..49S}
{Spoon}, H.~W.~W., {Marshall}, J.~A., {Houck}, J.~R., {Elitzur}, M., {Hao}, L.,
  {Armus}, L., {Brandl}, B.~R., \& {Charmandaris}, V. 2007, \apjl, 654, L49

\bibitem[{{Sturm} {et~al.}(2002){Sturm}, {Lutz}, {Verma}, {Netzer},
  {Sternberg}, {Moorwood}, {Oliva}, \& {Genzel}}]{2002A&A...393..821S}
{Sturm}, E., {Lutz}, D., {Verma}, A., {Netzer}, H., {Sternberg}, A.,
  {Moorwood}, A.~F.~M., {Oliva}, E., \& {Genzel}, R. 2002, \aap, 393, 821

\bibitem[{{Sturm} {et~al.}(2006){Sturm}, {Rupke}, {Contursi}, {Kim}, {Lutz},
  {Netzer}, {Veilleux}, {Genzel}, {Lehnert}, {Tacconi}, {Maoz}, {Mazzarella},
  {Lord}, {Sanders}, \& {Sternberg}}]{2006ApJ...653L..13S}
{Sturm}, E., \etal 2006, \apjl, 653, L13

\bibitem[{{Tommasin} {et~al.}(2008){Tommasin}, {Spinoglio}, {Malkan}, {Smith},
  {Gonz{\'a}lez-Alfonso}, \& {Charmandaris}}]{2008ApJ...676..836T}
{Tommasin}, S., {Spinoglio}, L., {Malkan}, M.~A., {Smith}, H.,
  {Gonz{\'a}lez-Alfonso}, E., \& {Charmandaris}, V. 2008, \apj, 676, 836

\bibitem[{{Tran}(2001)}]{2001ApJ...554L..19T}
{Tran}, H.~D. 2001, \apjl, 554, L19

\bibitem[{{Tran}(2003)}]{2003ApJ...583..632T}
---. 2003, \apj, 583, 632

\bibitem[{{Veilleux} \& {Osterbrock}(1987)}]{1987ApJS...63..295V}
{Veilleux}, S., \& {Osterbrock}, D.~E. 1987, \apjs, 63, 295

\bibitem[{{Verma} {et~al.}(2005){Verma}, {Charmandaris}, {Klaas}, {Lutz}, \&
  {Haas}}]{2005SSRv..119..355V}
{Verma}, A., {Charmandaris}, V., {Klaas}, U., {Lutz}, D., \& {Haas}, M. 2005,
  Space Science Reviews, 119, 355

\bibitem[{{V{\'e}ron-Cetty} \& {V{\'e}ron}(2006)}]{2006A&A...455..773V}
{V{\'e}ron-Cetty}, M.-P., \& {V{\'e}ron}, P. 2006, \aap, 455, 773

\bibitem[{{Weedman} {et~al.}(2005){Weedman}, {Hao}, {Higdon}, {Devost}, {Wu},
  {Charmandaris}, {Brandl}, {Bass}, \& {Houck}}]{2005ApJ...633..706W}
{Weedman}, D.~W., \etal 2005, \apj, 633, 706

\end{thebibliography}
\end{document}